\documentclass[10pt]{article}

\usepackage{amsmath}
\usepackage[round]{natbib}

\usepackage{hyperref}
\hypersetup{colorlinks=true,
linkcolor=black,          
   citecolor=black,        
    urlcolor=blue}
    
\usepackage{amsmath}

\usepackage{enumitem}

\usepackage{setspace}
\usepackage[table]{xcolor}
\usepackage{tocloft}
\usepackage{relsize}
\usepackage{epsfig}
\usepackage{graphics}
\usepackage{color}
\usepackage{longtable}
\usepackage[normalem]{ulem}
\usepackage[section]{placeins}
\usepackage{hanging}
\usepackage{epstopdf}
\usepackage{amssymb}
\usepackage{color}	    
\usepackage{rotating}		
\usepackage{subcaption}		
\usepackage{multirow}
\usepackage{booktabs}
\usepackage{adjustbox}
\usepackage[title]{appendix}
\usepackage{nameref}
\usepackage{cleveref}

\usepackage{xr}
\makeatletter
\newcommand*{\addFileDependency}[1]{
  \typeout{(#1)}
  \@addtofilelist{#1}
  \IfFileExists{#1}{}{\typeout{No file #1.}}
}
\makeatother

\newcommand*{\myexternaldocument}[1]{%
    \externaldocument{#1}%
    \addFileDependency{#1.tex}%
    \addFileDependency{#1.aux}%
}
\myexternaldocument{LCM2_appendices}

\graphicspath{{Figures/}}

\newcommand{\bs}{\begin{slide}}
\newcommand{\es}{\end{slide}}
\newcommand{\bc}{\begin{center}}
\newcommand{\ec}{\end{center}}
\newcommand{\bn}{\begin{enumerate}}
\newcommand{\en}{\end{enumerate}}
\newcommand{\bi}{\begin{itemize}}
\newcommand{\ei}{\end{itemize}}
\newcommand{\be}{\begin{eqnarray}}
\newcommand{\ee}{\end{eqnarray}}
\newcommand{\bes}{\begin{eqnarray*}}
\newcommand{\ees}{\end{eqnarray*}}

\newcommand{\ssection}[1]{\section[#1]{\centering\normalfont\scshape #1}}

\begin{document}

\title{Improving inference for nonlinear state-space models of animal population dynamics given biased sequential life stage data}
\maketitle
\date{}
\centerline{Leo Polansky\textsuperscript{1}, Ken B. Newman\textsuperscript{2,3}, Lara Mitchell\textsuperscript{2}}

Leo Polansky   \\
Corresponding author  \\
$^1$US Fish and Wildlife Service, San Francisco Bay-Delta Fish and Wildlife Office, Sacramento, CA, USA  \\
leo\_polansky@fws.gov  \\

Ken B. Newman   \\
$^2$US Fish and Wildlife Service, Lodi Fish and Wildlife Office, Lodi, CA, USA  \\
$^3$Current affiliation: Biomathematics \& Statistics Scotland and School of Mathematics, The University of Edinburgh, Scotland, UK  \\
ken.newman@bioss.ac.uk  \\

Lara Mitchell   \\
$^2$US Fish and Wildlife Service, Lodi Fish and Wildlife Office, Lodi, CA, USA  \\
lara\_mitchell@fws.gov  \\

\begin{abstract}
State-space models (SSMs) are a popular tool for modeling animal abundances. Inference difficulties for simple linear SSMs are well known, particularly in relation to simultaneous estimation of process and observation variances. Several remedies to overcome estimation problems have been studied for relatively simple SSMs, but whether these challenges and proposed remedies apply for nonlinear stage-structured SSMs, an important class of ecological models, is less well understood. Here we identify improvements for inference about nonlinear stage-structured SSMs fit with biased sequential life stage data. Theoretical analyses indicate parameter identifiability requires covariates in the state processes. Simulation studies show that plugging in externally estimated observation variances, as opposed to jointly estimating them with other parameters, reduces bias and standard error of estimates. In contrast to previous results for simple linear SSMs, strong confounding between jointly estimated process and observation variance parameters was not found in the models explored here. However, when observation variance was also estimated in the motivating case study, the resulting process variance estimates were implausibly low (near-zero). As SSMs are used in increasingly complex ways, understanding when inference can be expected to be successful, and what aids it, becomes more important. Our study illustrates (i) the need for relevant process covariates and (ii) the benefits of using externally estimated observation variances for inference for nonlinear stage-structured SSMs.
\end{abstract}


\maketitle

\section{Introduction}
\label{sec:introduction}
Understanding what determines changes in animal abundance through time is a basic question of population ecology and natural resource management. This in turn requires estimates of recruitment and survival, which can be used to quantify where, when, and how populations respond to endogenous (density-dependent) and exogeneous (external, density-independent) factors \citep{Turchin-2003}. However, fitting models can be difficult in practice because field survey-based estimates of animal population abundances are often very noisy and can be systematically biased in ways that are hard to correct for. For example, even after correcting for sampling inefficiencies by upwardly adjusting fish counts to reflect lower retention probabilities, and downwardly adjusting effort to reflect sampling efforts outside of the presumed habitat boundary, \cite*{Polansky-etal-DS-abund-2019} found that empirical survival ratios of juvenile survival were often biologically implausible. These implausible ratios, which suggest relative bias in sequential abundance estimates, and the need to understand the effects of covariates on population dynamics motivate the work here.

Direct analysis of the animal abundance time series data without consideration of the fact that variation in the series is a reflection of both observation (or measurement) errors, including sampling variation, as well as 
 natural process variation, which can include environmental and 
 demographic stochasticity, can yield misleading results. 
 The effects of ignoring observation error, thus assuming all variation 
 is process variation, have been examined for different population dynamic models; e.g., \citet{deValpine-Hastings-2002} showed
 biases in productivity rates in Beverton-Holt and Ricker models in a classical setting and \citet{Calder-etal-2003} showed bias in the
 productivity rate in a Beverton-Holt model  in a Bayesian setting.
Data with observation error can also lead to spurious detection of density dependent effects \citep{Freckleton-etal-2006}, the detectability of which can be a function of the strength of density dependence relative to the magnitude of observation error \citep{Knape-2008}. The inclusion of observation bias has been much less studied because most analyses use annual time step models of data based on a single survey. State-space models offer an attractive framework to model such data.

 State-space models (SSMs) \citep{Durbin-Koopman-2012, Newman-etal-2014}
 are models for two time series running in parallel. One series,
 the state process time series (hereafter the process time series), consists of values that may be viewed as an
 underlying description of the true state of a dynamic system.
 The true states are often unobservable, and
  sometimes called latent or hidden states \citep{Newman-etal-2006}. The other series, the observation time series, consists of known measurements on, or samples of, the state process. Measurements can be incorrect due to randomness in sampling as well as systematic biases \citep{Staudenmayer-Buonaccorsi-2006}. 
  The explicit separation of process variation from observation error 
 makes SSMs an attractive approach to analyzing ecological time
 series data. Relatedly, there is a divide-and-conquer appeal
 to SSMs in that one can formulate separate models aimed
 at explaining process dynamics and models that characterize
 imperfections in the observation, measurement, or estimation processes, which can be considerable in ecological data.
 The latter, namely the observation model, can be viewed almost as a 
 nuisance in the sense that
 even if one had perfect information of the process time series,
 explaining the dynamics of that alone can be quite challenging.

 However, one problem long recognized with some SSMs, especially  the
 special case of normal dynamic linear models (NDLMs; \citet{West-Harrison-2006}), which has been pointed out
 by several \citep{Dennis-etal-2006, Knape-2008,Knape-etal-2011, 
 AugerMethe-etal-2016}, is the difficulty of accurately estimating 
 both  process variance and observation variance. 
 \citet{Dennis-etal-2006} noted flatness in the profile likelihoods
 for the two variances in a simple univariate NDLM as well
 as a bimodal likelihood surface where a large process variance
 and small observation variance were as likely as a small process
 variance and a large observation variance. Another problem
 with NDLMs is overestimation of process variation when the
 ratio of observation variance to process variance is large
 \citep{AugerMethe-etal-2016}. There are similar difficulties in jointly estimating process variability and observation noise variances in nonlinear population dynamics process models as well \citep{deValpine-Hilborn-2005}.

Several remedies for these problems have been explored. One is to formulate the SSM such that the parameters of the
 process variance are also parameters for the mean(s) of
 the state process. Examples of such parameterizations with NDLMs include \citet{Newman-1998} who used normal approximations for
 binomial and multinomial distributions 
 where the expectations and variances matched those distributions,
 and similarly \citet{Besbeas-etal-2002} approximated Poisson and
 binomial distributions with the corresponding means and variances. A second remedy is to collect replicate observations that allow
 separation of observation variance from process variance
 \citep*{Dennis-etal-2010}. A third remedy is to avoid trying
 to estimate the observation variance altogether by inserting
 external estimates of observation variance when such estimates
 can be made \citep*{Knape-etal-2013}, e.g., variances of abundance estimates based on mark-recapture sampling. 
 
 With few exceptions, e.g., \citet{Knape-etal-2013}, most
 assessments of the problems of identifiability have been based
 on NDLMs and often for situations where the observations are
 annual.  In this paper we examine the estimation problem of a SSM whose state process is stage-structured, nonlinear, and non-Gaussian, and whose observations have stage-specific relative biases. Stage-structured population models are an important tool to draw inference about factors affecting recruitment and stage-specific survival rates \citep{Caswell-2001}. While some examples of their application embedded within a SSM inferential framework exist  \citep{deValpine2003,deValpine-Rosenheim-2008}, the effects of both noisy and biased data on inference for nonlinear SSMs are relatively poorly understood. Additionally, in practice ecologists are equally concerned with estimation of coefficients relating predictor variables to vital rates, also a topic of little focus compared to emphasis on estimation of variance parameters and latent states.
 
The rest of the paper is structured as follows. In Section~\ref{sec:model-description} we develop a stage-structured SSM that contains stochastic vital rates with covariate dependency and an observation model with bias terms for some life stages. We also describe several alternate formulations for the observation model depending on whether and how external estimates of observation variance are available. Section~\ref{sec:theoretical-inference} provides a theoretical analysis
of the identifiability of the parameters
for this model. Section~\ref{sec:simulation-study} presents a simulation study aimed to study parameter inference in practice, with a focus on estimates of latent states, state process parameters, and observation bias parameters. Attention is given to how estimation properties differ when fitting models that fix observation error variance using externally derived values compared to fitting models that simultaneously estimate observation error variance. Section~\ref{sec:case-study} describes a case study about delta smelt (\textit{Hypomesus transpacificus}) to illustrate the practical issues arising when fitting models of this type. Discussion is in Section~\ref{sec:discussion}.

\section{Model description}\label{sec:model-description}
\subsection{State process model}\label{sec:state-process-model}
We consider a population that can be partitioned into 
different life stages where  $n_{s,t}$ denotes the true abundance of life stage $s$ at time $t$. In particular we consider a fish population that
has four life stages: postlarvae, juveniles, sub-adults, and adults.
Given an initial abundance of reproducing adults $n_{A,t=0}$ in cohort $t=0$, the state process update equations are: 
\begin{align}
    \mbox{Postlarvae$|$Adults } &: 
       n_{PL,t}|n_{A,t-1} = \rho_t n_{A,t-1}\label{eqn:state-process-R} \\ 
    \mbox{Juveniles$|$Post-larvae } &: n_{J,t}|n_{PL,t} = 
      \phi_{PL,t} n_{PL,t}\label{eqn:state-process-phiPL} \\ 
    \mbox{Sub-adults$|$Juveniles } &: n_{SA,t}|n_{J,t} = \phi_{J,t} n_{J,t}\label{eqn:state-process-phiJ} \\ 
 \mbox{Adults$|$Sub-adults } &: n_{A,t}|n_{SA,t} = \phi_{SA,t} n_{SA,t}\label{eqn:state-process-phiSA}
\end{align}
where $\rho_{t}$ is time-specific recruitment and $\phi_{s,t}$ are the life-stage
and time-specific
survival rates. With (environmental) stochasticity in 4 processes, the vital
rate models for recruitment and the three survival rates are
  \begin{align}
\mbox{Recruitment } &: \rho_t \sim \mbox{LogNormal} \left(\textbf{x}^{T}_{R,t}\boldsymbol{\zeta},\sigma^2_{P,R}\right) \label{eqn:vital-R}\\
\mbox{Postlarval Survival } &: \phi_{PL,t} \sim \mbox{LogitNormal} \left(\textbf{x}^{T}_{PL,t}\boldsymbol{\beta},\sigma^2_{P,PL}\right)\label{eqn:vital-phiPL} \\
\mbox{Juvenile Survival } &: \phi_{J,t} \sim \mbox{LogitNormal} \left(\textbf{x}^{T}_{J,t}\boldsymbol{\eta},\sigma^2_{P,J}\right)\label{eqn:vital-phiJ} \\
\mbox{Sub-adult Survival } &: \phi_{SA,t} \sim \mbox{LogitNormal} \left(\textbf{x}^{T}_{SA,t}\boldsymbol{\gamma},\sigma^2_{P,SA}\right) \label{eqn:vital-phiSA}
\end{align}  
where $\boldsymbol{\zeta}^{T}=\left(\zeta_0,...,\zeta_{m_{R}}\right)$ is a
 vector of $m_{R}+1$ regression coefficients corresponding to a
 vector of recruitment predictor variables,
$\textbf{X}_{R,t}^{T}=(1,x_{1,t},...,X_{m_{R},t})$, and
$\sigma_{P,R}^2$ is the recruitment process variance on the log scale. The terms
in the survival functions, \Cref{eqn:vital-phiPL,eqn:vital-phiJ,eqn:vital-phiSA},
are defined analogously but with possibly different dimensions reflecting life
stage-specific numbers of covariates used in survival predictions.
The distribution designation of LogitNormal means that the logit transformed
survival probabilities are normally distributed.

\subsection{Observation model}
\label{sec:observation-model}
Observation error in the abundance estimates $\hat{n}_{s,t}$ can include
both bias $\psi_{s,t}$ so that $E[\hat{n}_{s,t} | n_{i,t}]=\psi_{s,t}n_{s,t}$, and sampling variance $V[\hat{n}_{s,t}]$ \citep{Staudenmayer-Buonaccorsi-2006}. Different choices about the conditional distribution for $\hat{n}_{i,t} | n_{i,t}$ can be made, which in turn can affect inference \citep{Knape-etal-2011}. We chose a lognormal distribution because it ensures strictly positive abundance indices (especially useful when abundance indices are near zero), it is perhaps the most common assumption (although more often in terms of a normal distribution when working with log abundances), and because prior work \citep{Polansky-etal-DS-abund-2019} suggests it to be preferred over a normal distribution for the case study model described in Section~\ref{sec:case-study}. 

Due to non-identifiability problems, we do not consider estimating biases that are both life stage and cohort specific because in that case the parameters would confound with the latent abundance estimates. Only life stage and survey type biases are allowed.  Motivated by the case study in Section \ref{sec:case-study}, we assume juvenile and sub-adult abundance indices are biased by factors $\psi_{J}$ and $\psi_{SA}$, and we use the following parameterizations for the observation model equations:  
\begin{align}
\hat{n}_{PL,t} &\sim 
  \mbox{Lognormal} \left ( \ln(n_{PL,t})-\sigma_{PL,t}^2/2, \sigma_{O,PL,t}^2 \right)\label{eqn:obs-model-PL}\\
 \hat{n}_{J,t} &\sim 
  \mbox{Lognormal} \left (  \ln(\psi_{J}n_{J,t})-\sigma_{J,t}^2/2, \sigma_{O,J,t}^2\right) \label{eqn:obs-model-J}\\
\hat{n}_{SA,t} &\sim 
  \mbox{Lognormal} \left ( \ln(\psi_{SA}n_{SA,t})-\sigma_{SA,t}^2/2, \sigma_{O,SA,t}^2 \right) \label{eqn:obs-model-SA}\\
\hat{n}_{A,t} &\sim 
  \mbox{Lognormal} \left ( \ln(n_{A,t})-\sigma_{A,t}^2/2, \sigma^2_{O,A,t} \right )\label{eqn:obs-model-A}  
\end{align}

With these parameterizations, the expected index is
\begin{equation}
E[\hat{n}_{s,t}]=\begin{cases}
n_{s,t} & \text{Observation bias absent}\\
\psi_{s}n_{s,t} & \text{Observation bias present}
\end{cases}
\label{eqn:obs-model-expected}
\end{equation}
and the variance is
\begin{equation}
V[\hat{n}_{s,t}]=\begin{cases}
(e^{\sigma_{O,s,t}^2}-1)n_{s,t}^2 & \text{Observation bias absent}\\
(e^{\sigma_{O,s,t}^2}-1)(\psi_{s}n_{s,t})^2 & \text{Observation bias present}
\end{cases}
\label{eqn:obs-model-variance}
\end{equation}
Irrespective of whether a bias factor is included, the coefficient of variation (CV) is
\begin{equation}
CV[\hat{n}_{s,t}]=\sqrt{e^{\sigma_{O,s,t}^2}-1}
\label{eqn:obs-model-CV}
\end{equation}
from which it follows that 
\begin{equation}
\sigma^2_{O,s,t}=\ln(CV[\hat{n}_{s,t}]^2+1)
\label{eqn:obs-model-sigma}
\end{equation}
Plugging the right hand side of \Cref{eqn:obs-model-sigma} into \Cref{eqn:obs-model-PL,eqn:obs-model-J,eqn:obs-model-SA,eqn:obs-model-A}, the observation model equations can be written as 
\begin{equation}
\hat{n}_{s,t} \sim 
\text{Lognormal}\left(\ln\left(\frac{\psi_{s}n_{s,t}}{\sqrt{CV[\hat{n}_{s,t}]^2+1}}\right),\ln(CV[\hat{n}_{s,t}]^2+1) \right)
\label{eqn:obs-model-equations-CV}
\end{equation}
where $\psi_{s}=1$ if there is no bias term present in a given life stage or survey specific observation model.

\subsection{Observation model formulations}
\label{sec:observation-model-formulations}
If along with point abundance estimates there are also available estimates of the variance of $\hat{n}_{s,t}$ derived \textit{externally} to the SSM, $\widehat{V}[\hat{n}_{s,t}]_{Ex}$, so that external estimates of the coefficient of variation
\begin{equation}
\widehat{CV}[\hat{n}_{s,t}]_{Ex}=\frac{\sqrt{\widehat{V}[\hat{n}_{s,t}]_{Ex}}}{\hat{n}_{s,t}}
\label{eqn:obs-model-CV-est} 
\end{equation}
and log-variance parameter
\begin{equation}
\hat{\sigma}^2_{O,s,t,Ex}=\ln\left(\widehat{CV}[\hat{n}_{s,t}]_{Ex}+1\right)
\label{eqn:obs-model-sigma-est} 
\end{equation}
can be obtained, three scenarios related to how the observation model equations are formulated for inference purposes are possible. In what follows the bias terms are included for generality with the understanding that $\psi_{s}=1$ if there is no bias term present in the life stage or survey specific observation model.

\begin{description}
\item[Scenario 1]- Use the external coefficient of variation estimates within \Cref{eqn:obs-model-equations-CV}. The observation model equations are of the form
\begin{equation}
\hat{n}_{s,t} \sim 
\text{Lognormal}\left(\ln\left(\frac{\psi_{s}n_{s,t}}{\sqrt{\widehat{CV}[\hat{n}_{s,t}]_{Ex}^2+1}}\right),\ln(\widehat{CV}[\hat{n}_{s,t}]_{Ex}^2+1) \right)
\label{eqn:obs-model-equations-CV-fixed}
\end{equation}

\item[Scenario 2]- Use the external abundance index variances $\widehat{V}[\hat{n}_{s,t}]_{Ex}$ but replace the denominator in \Cref{eqn:obs-model-CV-est} with the model estimated latent abundance value $n_{s,t}$, corrected with the estimated bias $\psi_{s}$. The observation models are of the form
\begin{equation}
\hat{n}_{s,t} \sim 
\text{Lognormal}\left(\ln\left(\frac{\psi_{s}n_{s,t}}{\sqrt{\widehat{V}[\hat{n}_{s,t}]_{Ex}/(\psi_{s}n_{s,t})^2+1}}\right),\ln(\widehat{V}[\hat{n}_{s,t}]_{Ex}/(\psi_{s}n_{s,t})^2+1) \right)
\label{eqn:obs-model-equations-sigmaOE-fixed}
\end{equation}

\item[Scenario 3]- Do not use external information about the variance of $\hat{n}_{s,t}$, and instead internally estimate observation variance along with the other parameters and latent states of the SSM. In this case, it would not be feasible to estimate both life stage and cohort specific values, but life stage-specific (and possibly survey method specific if multiple surveys are used to monitor a single life stage) parameters may be estimable. In this scenario, the observation model equations are similar to equations (\ref{eqn:obs-model-PL})-(\ref{eqn:obs-model-A}) but with a life stage $s$ specific observation log-variance parameter,
\begin{equation*}
\hat{n}_{s,t} \sim 
\text{Lognormal}\left (  \ln(\psi_{s}n_{s,t})-\sigma_{O,s,In}^2/2,\sigma_{O,s,In}^2\right)
\label{eqn:obs-model-equations-sigmaOE-est}
\end{equation*}
\end{description}
The choice between Scenario 1 and Scenario 2 may depend on whether the external coefficient of variation estimates $\widehat{CV}[\hat{n}_{s,t}]_{Ex}$ are better than the external observation variance estimates $\widehat{V}[\hat{n}_{s,t}]_{Ex}$.

\section{Theoretical identifiability}\label{sec:theoretical-inference}
We applied methods developed by \citet{Cole-McCrea-2016} to evaluate the
theoretical identifiability of the parameters in the SSM described in Section~\ref{sec:model-description} under various scenarios involving fixed or estimated observation error CV and the inclusion or exclusion
of covariates  in the state process model. An exhaustive summary vector based on expected values and variances of the observations was derived assuming each state process had an intercept and either zero or one covariate. The derivative matrix, formed by taking the partials of this vector with respect to the parameters of the model, was computed using \texttt{Maple~17} \citep{Maple-2013} and  code from the appendices of \citet{Cole-McCrea-2016}. If the rank of the derivative matrix, $r$, equals the number of parameters,
$p$, then all the model parameters are separately identifiable. 
If $r<p$, then $r$ parameters or parameter combinations are identifiable,
and \texttt{Maple} code from \citet{Cole-McCrea-2016} specifies  
which single parameters and combinations of parameters are identifiable.
 
One analysis examined the identifiability of parameters in the SSM defined by
 equations (\ref{eqn:state-process-R})-(\ref{eqn:obs-model-A}) with
 the restriction that a single covariate was used in the modeling of each 
 of the four state
 processes, namely, $x_{R,1,t}$ for recruitment, $x_{PL,1,t}$ for postlarval survival, $x_{J,1,t}$ for juvenile survival, and $x_{SA,1,t}$ for sub-adult
 survival.  The   model had
 18 parameters: 
 eight mean related parameters of the state process model ($\eta_0$,
 $\eta_1$, $\beta_0$, $\beta_1$, $\eta_0$, $\eta_1$, $\gamma_0$, $\gamma_1$), two mean related parameters for the observation model ($\psi_J$,
 $\psi_{SA}$), four variance parameters for the process model
 ($\sigma^2_{P,\rho}$,$\sigma^2_{P,PL}$, $\sigma^2_{P,J}$, $\sigma^2_{P,SA}$),
 and four variance parameters for the observation model assuming that these
 parameters were time-invariant ($\sigma^2_{O,PL}$, $\sigma^2_{O,J}$,
 $\sigma^2_{O,SA}$, $\sigma^2_{O,A}$). Crucially, the initial state
 component, $n_{A,0}$, was viewed as a known parameter, which seemed 
 reasonable given the assumption of unbiased estimates of the
 abundance of adults in the year prior to the start of the time series.
 An exhaustive summary vector of length 20 (which proved to be of 
 sufficient length) was constructed based on first order approximations
 of the expected values and variances of the observations. 
 A subset of the vector including examples of the expectations and
 variances is shown below.
 \begin{align*}
 E[y_{PL,1}] &= E[n_{PL,1}] \approx e^{\zeta_0+\zeta_1 x_{R,1}}n_{A,0}\\
 E[y_{J,1}] &= E[\psi_J n_{J,1}] \approx 
 \psi_J \frac{e^{\beta_0 + \beta_1 x_{PL,1}}}
 {1+e^{\beta_0 + \beta_1 x_{PL,1}}} e^{\zeta_0+\zeta_1 x_{R,1}}n_{A,0} \\
 E[y_{SA,1}] &= E[\psi_{SA} n_{SA,1}] \approx 
 \psi_{SA} \frac{e^{\eta_0 + \eta_1 x_{J,1}}}
 {1+e^{\eta_0 + \eta_1 x_{J,1}}}
 \frac{e^{\beta_0 + \beta_1 x_{PL,1}}}
 {1+e^{\beta_0 + \beta_1 x_{PL,1}}} e^{\zeta_0+\zeta_1 x_{R,1}}n_{A,0} \\
      V[y_{PL,1}] &= \sigma^2_{o,PL} + V[\rho_1 n_{A,0}]  \approx
      \sigma^2_{o,PL} +  n_{A,0}^2 \left ( e^{\zeta_0+\zeta_1 x_{R,1}} 
      \right )^2 e^{\sigma^2_{P,\rho}}-1
       \\
\end{align*}

The matrix of first derivatives of the exhaustive summary
vector with respect to each of the parameters had a full rank of 18,
thus all 18 parameters were identifiable.
Further details on the exhaustive summary calculations for this
SSM are provided in  \ref{app:model-identifiability}. 

A second set of analyses examined the effect of covariates for the
state processes on identifiability. This was largely motivated by
the fact that observations for two of the life stages in the case study (Section \ref{sec:case-study}) were biased. The potential for non-identifiability can be  seen
in the case of a SSM for a single life stage that is estimated with
bias and the state process dynamics are a simple auto-regression. For example, given $y_1$ $\approx$  $\psi n_1$ and $n_1$ $\approx$ 
$e^{\zeta_0} n_{0}$, then $E[y_1]$ $\approx$ $\psi e^{\zeta_0} n_{0}$ 
and difficulty separating $\psi$ and $\zeta_0$ is apparent.  The case 
for multiple life stages, where unbiased estimates were available for some life stages, is more complicated and the effect of including covariates 
for different processes in the state process model was of interest. Focusing 
solely on estimability of
mean parameters in the state and observation models, the effects on
identifiability of a single covariate being present or absent in each of 
the four process models ($2^4$=16 combinations) was examined.
In the case where the derivative matrix is not full rank, additional \texttt{Maple} code developed by \citet{Cole-McCrea-2016} determines which parameters
or combinations of parameters are identifiable. 

The results of this analysis are shown in Table 
\ref{tab:analytical.results}. In the most limited setting where no covariates
are used to model the process dynamics (case 1), there are six parameters,
but only four parameter combinations can be estimated, and the only
separately identifiable parameter is the intercept, $\zeta_0$, for
the process model dynamics. The identifiability of $\zeta_0$ is readily
seen given $n_{A,0}$ is assumed known and $y_{PL,1}$ is unbiased for
$n_{PL,1}$: $E[y_{PL,1}]$ $\approx$ $e^{\zeta_0}n_{A,0}$.  An example
of an identifiable combination, also from case 1, is $\psi_J\left(e^{\beta_0}/(1+e^{\beta_0})\right )$, which shows that
the bias in estimates of juvenile abundance cannot be separated from
the survival of post-larva to the juvenile stage.  When a covariate
is used for a given process, the intercept and slope parameters for
that process are always identifiable; e.g., in case 3 a covariate for
post-larval survival is included and $\beta_0$ and $\beta_1$ are separately
identifiable. When at least two processes have covariates (cases 9-16)
all the parameters are individually identifiable.  Whether or not the bias
parameters are identifiable is a function of both the inclusion of covariates
and the sequencing of the biases in the observations (here: unbiased,
biased, biased, and unbiased for post-larvae, juveniles, sub-adults,
and adults, respectively). For example  
parameter $\psi_J$ was identifiable when a covariate for the survival
of post-larva to the juvenile stage was included (e.g., case 3), while  
  $\psi_{SA}$ was identifiable when
a covariate for the survival of sub-adults to the adult stage was
included (e.g., case 5).  We note that additional analyses of the effects
of covariates and biases showed that if covariates were available for
all four processes, and observations for all four life stages were biased,
i.e., there were also $\psi_{PL}$ and $\psi_A$, say, all the parameters
were identifiable (e.g., cases 17-18).

\begin{table}[ht]
    \caption{Summary of parameter identifiability results for state process mean parameters and observation bias parameters conditional on the inclusion or exclusion of covariates in a given set of state processes.}
    \label{tab:analytical.results}
\centering
\begin{tabular}{>{\centering}p{.5cm}lp{7.52cm}>{\centering}p{1.85cm}l}
  \hline  \hline
     &                 &               & Number of     &  \\ 
     &                 &               & Identifiable  & Identifiable \\ 
     & State Processes &               & Parameter     & Singleton \\ 
Case & with Covariates & Parameters to be Estimated (total) & Combinations  & Parameters \\ 
  \hline
1 & None & \{$\zeta_0$,$\beta_0$,$\eta_0$,$\gamma_0$,$\psi_J$,$\psi_{SA}$\} (6) & 4 & $\zeta_0$ \\
2 & $\rho$ & \{$\zeta_0$,$\zeta_1$,$\beta_0$,$\eta_0$,$\gamma_0$,$\psi_J$,$\psi_{SA}$\} (7) & 5 & $\zeta_0$,$\zeta_1$ \\
3 & $\phi_{PL}$ & \{$\zeta_0$,$\beta_0$,$\beta_1$,$\eta_0$,$\gamma_0$,$\psi_J$,$\psi_{SA}$\} (7) & 6 & $\zeta_0$,$\beta_0$,$\beta_1$,$\psi_J$ \\
4 & $\phi_{J}$ & \{$\zeta_0$,$\beta_0$,$\eta_0$,$\eta_1$,$\gamma_0$,$\psi_J$,$\psi_{SA}$\} (7) & 6 & $\zeta_0$,$\eta_0$,$\eta_1$  \\
5 & $\phi_{SA}$ & \{$\zeta_0$,$\beta_0$,$\eta_0$,$\gamma_0$,$\gamma_1$,$\psi_J$,$\psi_{SA}$\} (7) & 6 & $\zeta_0$,$\gamma_0$,$\gamma_1$,$\psi_{SA}$ \\
6 & $\rho$,$\phi_{PL}$ & \{$\zeta_0$,$\zeta_1$,$\beta_0$,$\beta_1$,$\eta_0$,$\gamma_0$,$\psi_J$,$\psi_{SA}$\} (8) & 7 & $\zeta_0$,$\zeta_1$,$\beta_0$,$\beta_1$,$\psi_{J}$ \\
7 & $\rho$,$\phi_{J}$ & \{$\zeta_0$,$\zeta_1$,$\beta_0$,$\eta_0$,$\eta_1$,$\gamma_0$,$\psi_J$,$\psi_{SA}$\} (8) & 7 & $\zeta_0$,$\zeta_1$,$\eta_0$,$\eta_1$ \\
8 & $\rho$,$\phi_{SA}$ & \{$\zeta_0$,$\zeta_1$,$\beta_0$,$\eta_0$,$\gamma_0$,$\gamma_1$,$\psi_J$,$\psi_{SA}$\} (8) & 7 & $\zeta_0$,$\zeta_1$,$\gamma_0$,$\gamma_1$,$\psi_{SA}$ \\
9 & $\phi_{PL}$,$\phi_{J}$ & \{$\zeta_0$,$\beta_0$,$\beta_1$,$\eta_0$,$\eta_1$,$\gamma_0$,$\psi_J$,$\psi_{SA}$\} (8) & 8 & All \\
10 & $\phi_{PL}$,$\phi_{SA}$ & \{$\zeta_0$,$\beta_0$,$\beta_1$,$\eta_0$,$\gamma_0$,$\gamma_1$,$\psi_J$,$\psi_{SA}$\} (8) & 8 & All \\
11 & $\phi_{J}$,$\phi_{SA}$ & \{$\zeta_0$,$\beta_0$,$\eta_0$,$\eta_1$,$\gamma_0$,$\gamma_1$,$\psi_J$,$\psi_{SA}$\} (8) & 8 & All \\
12 & $\rho$,$\phi_{PL}$,$\phi_{J}$ & \{$\zeta_0$,$\zeta_1$,$\beta_0$,$\beta_1$,$\eta_0$,$\eta_1$,$\gamma_0$,$\psi_J$,$\psi_{SA}$\} (9) & 9 & All \\
13 & $\rho$,$\phi_{PL}$,$\phi_{SA}$ & \{$\zeta_0$,$\zeta_1$,$\beta_0$,$\beta_1$,$\eta_0$,$\gamma_0$,$\gamma_1$,$\psi_J$,$\psi_{SA}$\} (9) & 9 & All \\
14 & $\rho$,$\phi_{J}$,$\phi_{SA}$ & \{$\zeta_0$,$\zeta_1$,$\beta_0$,$\eta_0$,$\eta_1$,$\gamma_0$,$\gamma_1$,$\psi_J$,$\psi_{SA}$\} (9) & 9 & All \\
15 & $\phi_{PL}$,$\phi_{J}$,$\phi_{SA}$ & \{$\zeta_0$,$\beta_0$,$\beta_1$,$\eta_0$,$\eta_1$,$\gamma_0$,$\gamma_1$,$\psi_J$,$\psi_{SA}$\} (9)& 9 & All \\
16 & $\rho$,$\phi_{PL}$,$\phi_{J}$,$\phi_{SA}$ & \{$\zeta_0$,$\zeta_1$,$\beta_0$,$\beta_1$,$\eta_0$,$\eta_1$,$\gamma_0$,$\gamma_1$,$\psi_J$,$\psi_{SA}$\} (10) & 10 & All \\
 17 & $\rho$,$\phi_{PL}$,$\phi_{J}$,$\phi_{SA}$ & \{$\zeta_0$,$\zeta_1$,$\beta_0$,$\beta_1$,$\eta_0$,$\eta_1$,$\gamma_0$,$\gamma_1$,$\psi_J$,$\psi_J$,$\psi_{SA}$,$\psi_{PL}$,$\psi_A$\} (12) & 12 & All \\
 18 & $\rho$,$\phi_{J}$,$\phi_{SA}$ & \{$\zeta_0$,$\zeta_1$,$\beta_0$,$\beta_1$,$\eta_0$,$\eta_1$,$\gamma_0$,$\gamma_1$,$\psi_J$,$\psi_J$,$\psi_{SA}$,$\psi_{PL}$,$\psi_A$\} (11) & 11 & All \\
   \hline
\end{tabular}
\end{table}

These theoretical calculations are based on a frequentist model formulation, while in Sections \ref{sec:simulation-study} and \ref{sec:case-study} we apply Bayesian methods to fit the SSM. Non-identifiability in a frequentist context implies non-identifiability in a Bayesian context with uninformative priors, though it is possible for informative priors to help alleviate this issue \citep{Cole-McCrea-2016}. We note that Scenarios 1 and 2 from Section \ref{sec:observation-model-formulations} assume observation error CV or variance priors that are perfectly informative, i.e., the probability of the external estimate is one. 

\section{Simulation study}\label{sec:simulation-study}

\subsection{Design}

Although we identified conditions under which parameter identifiability is ensured theoretically, data-specific features can result in practically non-identifiable, or nearly redundant, parameters, sometimes identified by flat profile likelihoods and infinitely large confidence intervals in a
frequentist context \citep{Raue2009}. We therefore used simulations to explore practical identifiability and to explore the potential benefits of fixing external estimates of $CV[\hat{n}_{s,t}]_{Ex}$ as in Scenario 1. Data sets were generated in \texttt{R} \citep{R_cite-2019} according to equations~(\ref{eqn:state-process-R})-(\ref{eqn:obs-model-A}) with 20 cohorts. Motivated by the case study (Section \ref{sec:case-study}), the postlarval and adult life stages were assumed to have no observation bias ($\psi_{PL} = \psi_{A} = 1$) while the juvenile and sub-adult life stages had observation biases less than one. Each data set was used to fit two models, one with $CV[\hat{n}_{s,t}]$ fixed at externally-derived estimates (Scenario 1) and one with $CV[\hat{n}_{s,t}]$ internally estimated as part of the model (Scenario 3). In the first model, the fixed values were assumed to be potentially imperfect according to the distribution $\widehat{CV}[\hat{n}_{s,t}]_{Ex} \sim \text{Unif} \Big( (1 - a) CV[\hat{n}_{s,t}], (1 + a) CV[\hat{n}_{s,t}] \Big)$ where $CV[\hat{n}_{s,t}]$ is the true value and $a \in [0,1)$. 

We ran 100 simulations each for $a = 0$, in which case $\widehat{CV}[\hat{n}_{s,t}]_{Ex}$ is equal to the true value $CV[\hat{n}_{s,t}]$, and for $a = 0.5$, which represents a more realistic case where $\widehat{CV}[\hat{n}_{s,t}]_{Ex}$ is estimated imperfectly. True values of $CV[\hat{n}_{s,t}]$ were generated from a Uniform(0.1, 1)
distribution. The recruitment and three survival processes were each
functions of single covariates. To incorporate the concept of model selection in the study along with parameter estimation, we included two potential covariates per process (the true covariate used to generate data and a second covariate) in the fitted models. The true values used to generate data and the prior distributions used for model fitting are shown in \ref{app:sim-study} (Table~\ref{tab:sim.true.value.table}). Model fitting here and in Section~\ref{sec:case-study} used Bayesian methods implemented within \texttt{R} \citep{R_cite-2019} using \texttt{JAGS} v4.3.0 \citep{Plummer-2003,Plummer-2016,Su-Yajima-2015}. Model performance was evaluated by calculating marginal posterior summary statistics as well as relative bias, i.e., (posteriorMean - trueValue)/trueValue. Model convergence was assessed by examining trace plots for signs of poor mixing and calculating Gelman-Rubin statistics.

\subsection{Results}

\begin{figure}[h]
    \centering
    \includegraphics[scale=1]{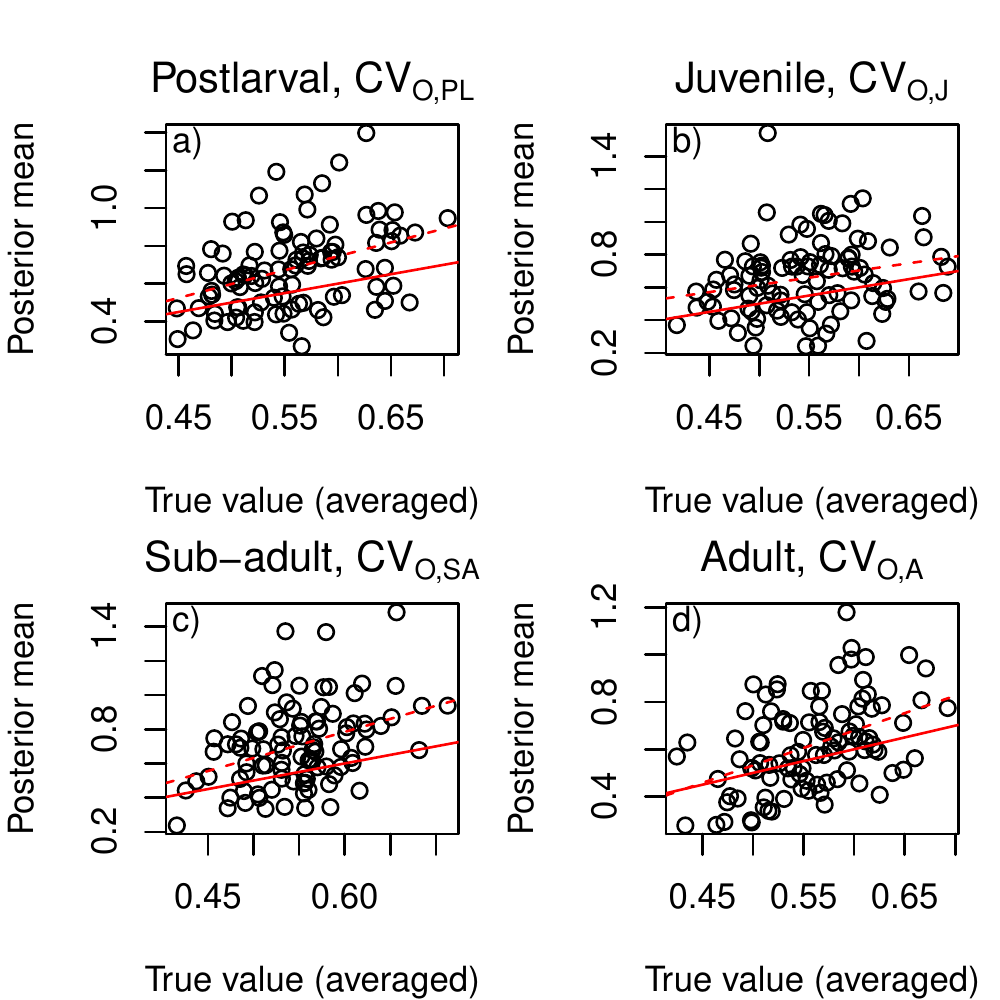}
    \caption{Observation error CV posterior means vs. true observation CV (averaged over cohort years) by life stage from the simulation study with $a=0.5$. One-to-one lines (solid) and fitted linear regression lines (dashed) are shown on each panel for reference.}
    \label{fig:sim.results.CVOE.scatterplots}
\end{figure}

\begin{figure}[h]
    \centering
    \includegraphics[scale=0.7]{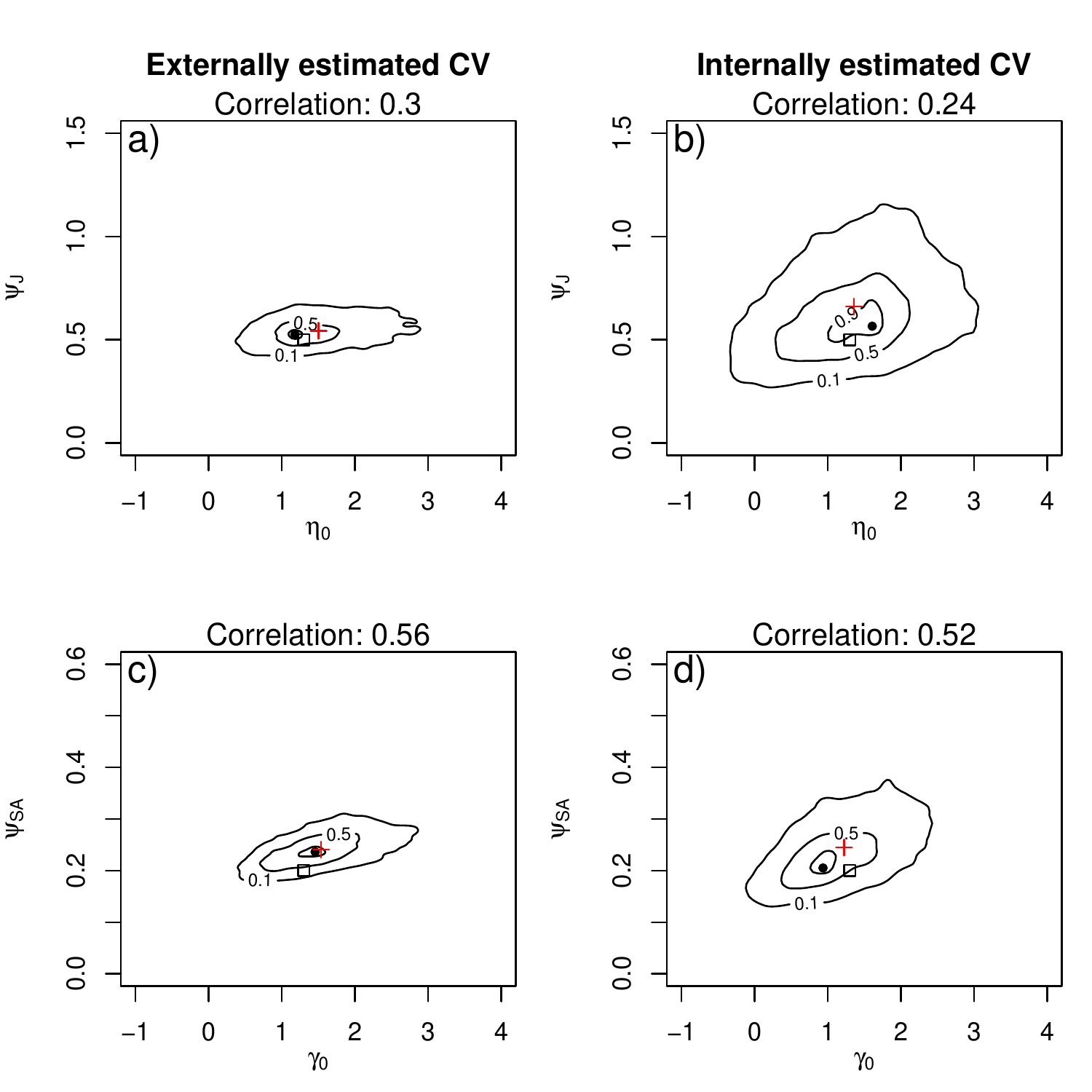}
    \caption{Joint posterior density plots of the observation bias parameter and survival intercept for juvenile (panels a and b) and sub-adult (panels c and d) survival sub-processes based on a single simulation with $a=0.5$. The left and right columns correspond to externally and internally estimated observation error CV. Joint density plots are scaled to lie between 0 and 1, levels are drawn at the 0.9, 0.5, and 0.1 quantiles, boxes reflect true parameter values, and crosses reflect marginal means.}
    \label{fig:sim.results.psi.intercept.jposteriors}
\end{figure}

\begin{figure}[h]
    \centering
    \includegraphics[scale=1]{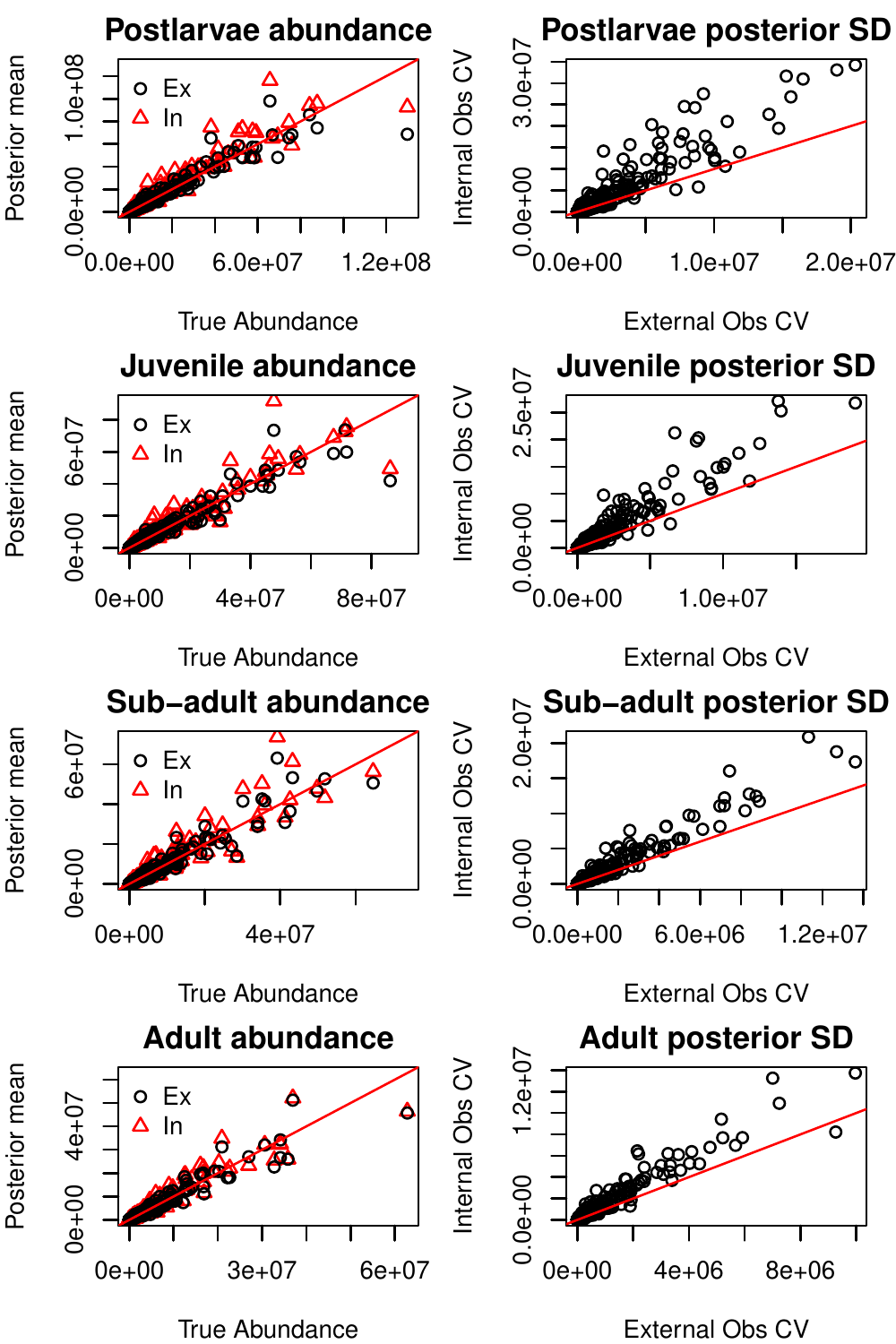}
    \caption{Comparison of postlarval, juvenile, sub-adult, and adult latent abundance posterior means and standard deviations (SD) from the simulation study with $a=0.5$. Left column shows posterior mean vs. true abundance for models with externally (Ex) and internally (In) estimated observation error CV. Right column shows posterior standard deviations from the model with internal observation error CV estimates (Internal Obs CV) against standard deviations from the model with external estimates (External Obs CV). A one-to-one line is shown in each panel for reference.}
    \label{fig:sim.results.abund.sd.scatterplot}
\end{figure}

Posterior means of the vital rate coefficients, process variance, and observation bias parameters, averaged across simulations, were similar whether observation error was externally or internally estimated and whether the level of noise in the external estimates was low ($a=0$) or high ($a=0.5$) (Table \ref{tab:sim.results.param}). On average, observation bias parameters were well-estimated, while recruitment process variance exhibited the highest level of relative bias (Table \ref{tab:sim.results.param}). Internally estimated observation CV was also generally biased high relative to the mean true observation CV for each life stage (Figure \ref{fig:sim.results.CVOE.scatterplots}). Plugging in external estimates of observation CV resulted in lower average posterior standard deviations and relative biases, as well as reduced diffusivity in joint posteriors; for example, see Figure \ref{fig:sim.results.psi.intercept.jposteriors}. Latent abundance posterior means were similarly well-estimated regardless of how observation CV was handled, though plugging in external observation CV estimates can lead to smaller abundance standard deviations (Figure \ref{fig:sim.results.abund.sd.scatterplot}). The findings presented here for the case with $a=0.5$ are qualitatively similar to those for the case with $a=0$.

\begin{sidewaystable}
 \caption{Summary of simulation study parameter estimates. Average posterior mean (Mean), posterior standard deviation (SD), and relative bias (Rel Bias) were calculated across simulations for the cases with $a=0$ and $a=0.5$. $M_{Ex}$ and $M_{In}$ represent the models with externally and internally estimated observation error, respectively.}
    \label{tab:sim.results.param}
     \begin{tabular}{lcrrrrrrp{0.001cm}rrrrrr}
     \hline \hline
      \centering
            &&\multicolumn{6}{c}{(a) Simulation with $a=0$.} && \multicolumn{6}{c}{(b) Simulation with $a=0.5$.}\\
              \cline{3-8}\cline{10-15}
             & True & \multicolumn{2}{c}{Mean} & \multicolumn{2}{c}{SD} & \multicolumn{2}{c}{Rel Bias} && \multicolumn{2}{c}{Mean} & \multicolumn{2}{c}{SD} & \multicolumn{2}{c}{Rel Bias} \\ 
             \cmidrule(lr){3-4}\cmidrule(lr){5-6}\cmidrule(lr){7-8}\cmidrule(lr){10-11}\cmidrule(lr){12-13}\cmidrule(lr){14-15}
            Parameter & Value & $M_{Ex}$ & $M_{In}$ & $M_{Ex}$ & $M_{In}$ & $M_{Ex}$ & $M_{In}$ &&$M_{Ex}$ & $M_{In}$ & $M_{Ex}$ & $M_{In}$ & $M_{Ex}$ & $M_{In}$ \\ 
              \hline
              $\zeta_0$ & 1.00 & 1.02 & 1.07 & 0.12 & 0.18 & 0.02 & 0.07 && 1.03 & 1.08 & 0.11 & 0.18 & 0.03 & 0.08 \\ 
              $\zeta_1$ & 1.00 & 0.98 & 0.97 & 0.09 & 0.12 & -0.02 & -0.03 && 0.99 & 1.00 & 0.10 & 0.12 & -0.01 & 0.00 \\ 
              $\zeta_2$ & 0.00 & 0.00 & 0.01 & 0.09 & 0.12 & NA & NA &&0.01 & 0.00 & 0.09 & 0.12 & NA & NA \\ 
              $\beta_0$ & 1.30 & 1.33 & 1.28 & 0.60 & 0.67 & 0.02 & -0.01 && 1.34 & 1.27 & 0.60 & 0.68 & 0.03 & -0.02 \\ 
              $\beta_1$ & 1.00 & 0.82 & 0.70 & 0.39 & 0.45 & -0.18 & -0.30 && 0.78 & 0.69 & 0.40 & 0.45 & -0.22 & -0.31 \\ 
              $\beta_2$ & 0.00 & 0.01 & 0.00 & 0.36 & 0.43 & NA & NA &&  0.00 & -0.02 & 0.37 & 0.43 & NA & NA \\ 
              $\eta_0$ & 1.30 & 1.34 & 1.32 & 0.60 & 0.67 & 0.03 & 0.01 && 1.32 & 1.28 & 0.60 & 0.66 & 0.02 & -0.01 \\ 
              $\eta_1$ & 1.00 & 0.80 & 0.68 & 0.40 & 0.46 & -0.20 & -0.32 && 0.82 & 0.71 & 0.40 & 0.45 & -0.18 & -0.29 \\ 
              $\eta_2$ & 0.00 & -0.01 & 0.02 & 0.36 & 0.43 & NA & NA &&  -0.05 & -0.04 & 0.36 & 0.43 & NA & NA \\ 
              $\gamma_0$ & 1.30 & 1.40 & 1.37 & 0.59 & 0.66 & 0.08 & 0.05 &&  1.46 & 1.43 & 0.60 & 0.67 & 0.12 & 0.10 \\ 
              $\gamma_1$ & 1.00 & 0.85 & 0.76 & 0.40 & 0.46 & -0.15 & -0.24 && 0.75 & 0.66 & 0.42 & 0.47 & -0.25 & -0.34 \\ 
              $\gamma_2$ & 0.00 & -0.01 & -0.05 & 0.37 & 0.43 & NA & NA && 0.02 & -0.01 & 0.39 & 0.45 & NA & NA \\ 
              $\sigma_{P,R}$ & 0.05 & 0.11 & 0.15 & 0.09 & 0.12 & 1.27 & 1.93 &&  0.15 & 0.15 & 0.10 & 0.11 & 2.04 & 2.01 \\ 
              $\sigma_{P,PL}$ & 0.50 & 0.48 & 0.53 & 0.36 & 0.43 & -0.04 & 0.06 && 0.58 & 0.56 & 0.41 & 0.44 & 0.16 & 0.12 \\ 
              $\sigma_{P,J}$ & 0.50 & 0.52 & 0.58 & 0.38 & 0.45 & 0.03 & 0.16 &&  0.53 & 0.55 & 0.39 & 0.43 & 0.07 & 0.10 \\ 
              $\sigma_{P,SA}$ & 0.50 & 0.51 & 0.55 & 0.38 & 0.44 & 0.01 & 0.10 && 0.59 & 0.61 & 0.42 & 0.46 & 0.18 & 0.21 \\ 
              $\psi_J$ & 0.50 & 0.52 & 0.55 & 0.08 & 0.15 & 0.04 & 0.10 && 0.51 & 0.53 & 0.08 & 0.14 & 0.03 & 0.07 \\ 
              $\psi_{SA}$ & 0.20 & 0.21 & 0.23 & 0.03 & 0.06 & 0.05 & 0.13 &&  0.21 & 0.23 & 0.03 & 0.06 & 0.05 & 0.15 \\
                  \hline
            \end{tabular}
\end{sidewaystable}

\section{Case study}\label{sec:case-study}
We used the model described in Section~\ref{sec:model-description} to quantify the relative importance of salient factors posited to determine the population dynamics of delta smelt (\textit{Hypomesus transpacificus}). Delta smelt are a small, nearly annual pelagic fish endemic to the interior ``Delta'' portion of the San Francisco Estuary whose life history is conceptually well understood \citep{Moyle-Herbold-1992, Moyle-etal-2016}. Spawning typically takes place during the winter and spring, with offspring maturing through a number of life stages to eventually become spawning adults by winter of the next calendar year.

Starting with the recruitment model \Cref{eqn:vital-R} that connected winter adult abundance to the next cohort's postlarval abundance in late spring, survival models for summer, fall and winter following \Cref{eqn:state-process-phiPL,eqn:state-process-phiJ,eqn:state-process-phiSA} connected the subsequent three life stages of juveniles, sub-adults, and adults, respectively. Abundance indices and coefficients of variation developed in \citet{Polansky-etal-DS-abund-2019} were used as observations on the abundances of each life stage starting with the 1994 birth cohort adult abundance index and including the subsequent 21 cohorts. Based on the abundance index construction work in \citet{Polansky-etal-DS-abund-2019}, we assumed postlarva indices, and adult abundance indices for $t>6$, were unbiased relative to an overall unknown scaling factor, i.e. $\psi_{PL}=1$ and $\psi_{A,t}=1$ for $t>6$. Juvenile and sub-adult observation biases, $\psi_{J}$ and $\psi_{SA}$, respectively, were estimated. Because the sampling method used to collect data about adult abundances prior to 2002 was the same as that used  for sub-adult data collection, and indices include length based corrections, we set $\psi_{A,t}=\psi_{SA}$ for $t \leq 6$. Covariate data used to model recruitment and survival consisted of a collection of abiotic habitat condition metrics (e.g. flow and temperature), abundance indices of bottom up (e.g. prey) and top down (e.g. predators) trophic drivers, and competitors, summarized in    \ref{app:case-study} Table~\ref{tab:predictor-variables}. Further details, data sets, and code are available from the corresponding author.

 The steps of assembling covariate sets to model each vital rate and the complete set of  parameter posterior results are described in detail in    \ref{app:case-study}. This included removal of highly correlated within life stage predictor variables (typically different hydrodynamic measures but also including summer inflow and food) and early life stage striped bass (\textit{Morone saxatilis}) on recruitment and survival because the predicted influence was opposite to what would be expected based on direct interspecific interaction considerations alone. The predicted vital rates for these covariates are shown in Figure~\ref{fig:set-of-focal-models}. Model validation was done in several ways, including graphical posterior checks of response residuals \citep{Gelman-Shalizi-2013}, and q-q plots of one step ahead forecast residuals \citep{Smith1985}. 

Two models with the same sets of covariates were fit, one using externally estimated observation error CV (Scenario 1), and one that estimated them internally (Scenario 3). The model based on Scenario 1 estimated lower process noise and higher observation error than the model using Scenario 3, and, as predicted by the simulation study, had increased posterior standard deviations for both latent abundances and observation bias. Posterior distributions of the bias parameters in both cases showed them to be considerably smaller than 1, with the $\psi_{J}$ being larger than $\psi_{SA}$.

We found a number of covariates with more support than others for each vital rate. Here we present a subset of these results to illustrate the most important vital rate predictions while relegating the full set of results and details to~\ref{app:case-study}.  Using the results from the model fit with external estimates of observation error CV plugged in and selecting (somewhat arbitrarily) a 0.80 value as an evidence threshold, recruitment was most influenced by temperature (negatively), the prior fall X2 location, and adult food; postlarval survival by inflow and turbidity; juvenile survival by turbidity and temperature (positively); and sub-adult survival by south Delta turbidity and hydrodynamics, adult food, and adult striped bass. Of the predator/competitor indices considered here, only the effect of summer inland silversides (\textit{Menidia beryllina}) abundance on postlarval survival and juvenile and adult striped bass on sub-adult survival had biologically plausible negative expected effects. 

\begin{figure}
    \centering
    \includegraphics[scale=1]{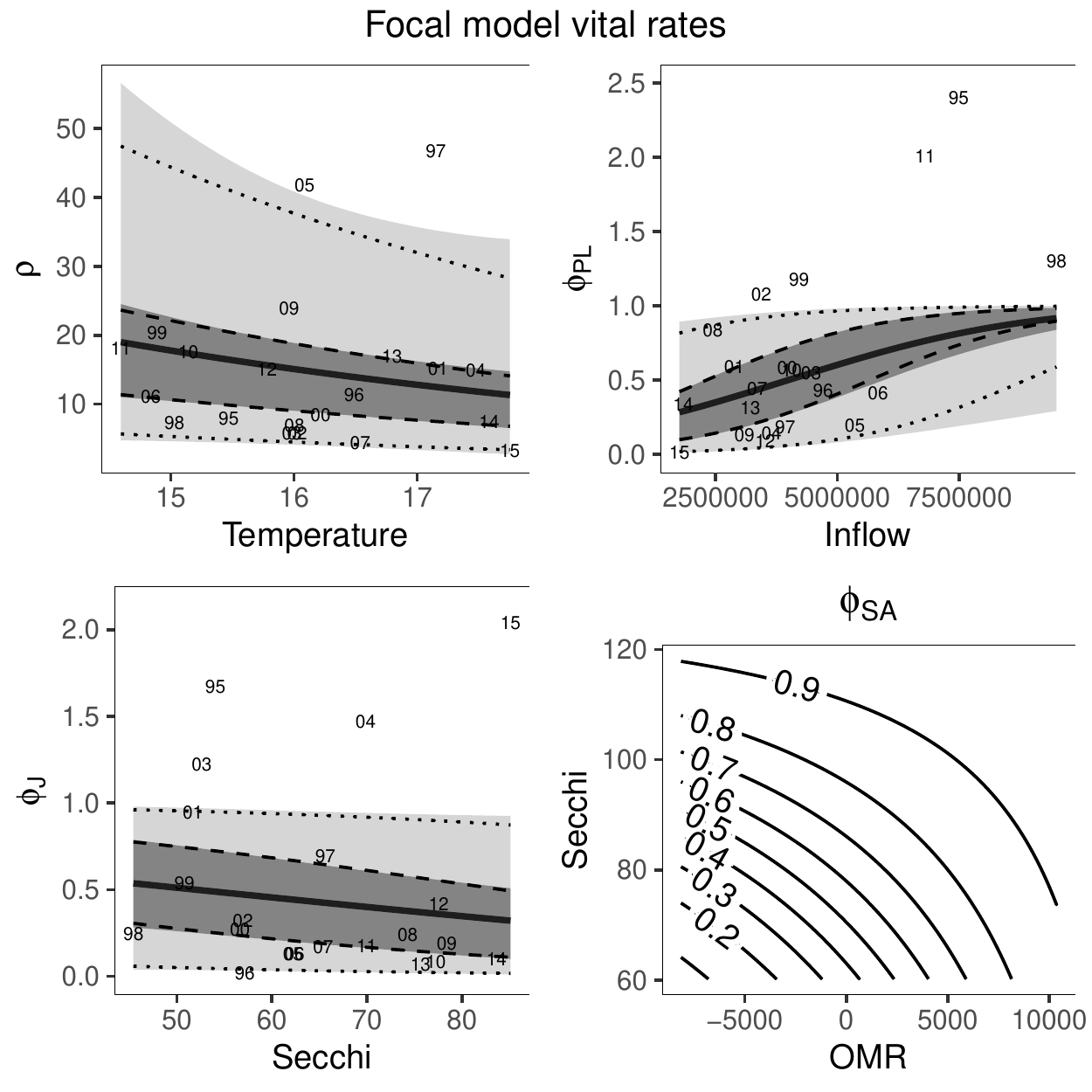}
    \caption{Predicted vital rates for the focal model. The solid curved lines show expected values, dark and light grey shadings show the $100(1-\alpha)\%$ central credible intervals for $\alpha=0.5$ and $\alpha=0.05$, respectively, and include posterior parameter estimate uncertainty. The dashed and dotted lines show the 50\% and 95\%, respectively, central credible intervals using the mean values of the posterior. Ratios of abundance indices are shown with the last two digits of the cohort year, adjusted by the inverse of the posterior means of observation error bias when relevant.}
    \label{fig:Focal_model_vital_rates}
\end{figure}

To illustrate vital rate predictions, a model including only the covariates with the highest posterior evidence for each vital rate was constructed. Also included was an interaction between the two covariates that most impact sub-adult survival, both of which had near one evidence of support. There is little difference between the distribution of vital rates when parameter estimate uncertainty is not included, and the interquartile prediction range is considerably more bounded than the 95\% prediction interval (Figure~\ref{fig:Focal_model_vital_rates}). Increases in sub-adult survival with decreases in turbidity (increases in South Secchi) become more pronounced as OMR decreases (Figure~\ref{fig:Focal_model_vital_rates}).

\section{Discussion}\label{sec:discussion}

 Identifiability issues for SSMs fit to ecological data sets have
 primarily concentrated on normal dynamic linear models  
 (\citet{deValpine-Hilborn-2005,deValpine-Rosenheim-2008} provide some exceptions), with specific focus on accurately distinguishing the magnitudes
 of process and observation variances \citep{Dennis-etal-2006,
 Dennis-etal-2010, Knape-etal-2011, AugerMethe-etal-2016}, and
 difficulties in 
 accurately assessing density dependence in SSMs for population
 dynamics \citep{Freckleton-etal-2006, Knape-2008}.  The effects of
 covariates  and the effects of external estimates of observation variance on identifiability, particularly for nonlinear and non-Gaussian SSMs, have not received as much attention.
 
For the nonlinear non-Gaussian SSMs examined here, the utility of covariates for enabling identifiability, particularly in the case of biased observations, was clear. Without covariates, but with bias in some of the observations, non-identifiability occurs. In some cases, one can use method of moments to determine identifiability by setting the observations to their expected values and solving the resulting system of equations. However, the algebra involved in this approach can be extremely difficult and determining what parameter combinations are identifiable can also be challenging. The recently developed methods of \citet{Cole-McCrea-2016} are attractive and elegant tools for analytically assessing parameter identifiability for SSMs in a far less algebraically tedious manner, indicating both how many parameter combinations are identifiable as well as what they are. These methods deserve routine application both to guide SSM formulation as well as for after-the-fact assessment. A cautionary note, however, is that theoretical identifiability does not rule out practical non-identifiability for a given data set \citep{Raue2009}.  Simulating data sets from a hypothesized SSM and then examining the ability to estimate the known parameters can be a helpful exercise for identifying practical estimation difficulties and data specific modeling challenges.

Based on the simulation study, the improvement in inference when observation error CV was externally estimated (Scenario 1) compared to when it was internally estimated (Scenario 3) was relatively minor. The primary advantage of using external estimates was an increase in precision, particularly for the observation bias and state estimates. As expected, these gains decrease when the external estimates of observation variance are themselves measured with error. Case study results based on these two scenarios mostly mirror the simulation study findings, particularly with respect to inference about latent states and observation bias parameters. However, an important difference is that when observation error CV was estimated internally estimation difficulties appeared to be greatly exacerbated in general, with complex joint posterior distributions containing multiple modes and ridges, particularly in the process variance and intercept parameter dimensions. One consequence was that process variance estimates were unrealistically low. The exact aspect of the case study data set responsible for the low process noise estimates is not clear.

A practical concern when modeling empirical data is that relationships between covariates and response variables may appear weaker than they are 
in reality because of noise in the covariate data, also known as error in variables \citep{Carroll-2006}, or because the covariate data were summarized in a non-optimal way \citep{Ferguson-2017}. Exploratory simulations indicated that covariate noise affects estimated vital rates in the SSM presented here, making this an aspect of the model that requires further development. An estimation challenge that seems particularly challenging given the model framework considered here is that as effect size diminishes, non-identifiability issues can emerge. Including covariates does not guarantee that straightforward application of computational methods will be sufficient for estimation.

The delta smelt population modeling presented here integrated data from more surveys than has been done previously. While the sequential life stage model required addressing relative bias in abundance estimates, this model framework allowed new insights about drivers of population abundances. For example, we found flow related impacts on summer survival and lagged fall flow effects on recruitment, whereas prior analyses by \citet{Feyrer-2011} were unable to precisely identify where in the life cycle the flow effects on population dynamics occur. Extending the findings by \citet{Grimaldo-2009}, who found that OMR predicts an index of south Delta mortality, we found this spatially localized hydrodynamics measure along with a spatially localized measure of water clarity was correlated with population wide sub-adult survival. Because inference about these factors was embedded within a population dynamics model, it is possible to carry out population viability analysis sensu \citet{Morris-Doak-2002}, an important model application. We also identified a number of predator/competitor relationships to delta smelt vital rates, some of which were biologically implausible, suggesting that some cohabitant species are more influenced in the contemporary Delta by shared habitat conditions than inter-specific interactions, although more complicated models remain to be explored.

\citet{AugerMethe-etal-2016} remarked that SSMs are ``becoming the favored statistical framework [in ecology]
for modeling animal movement and population dynamics.'' The state-process equations describing realistic population models are often nonlinear, non-Guassian, and link multiple life stages. Because they allow integration of multiple datasets, the SSM is an important tool for advancing population modeling in practice. Here we focused on a particular matrix model and arrangement of where observation bias enters the observation model. Matrix models linking the life stages with vital rates can encompass a wide variety of life history strategies \citep{Caswell-2001}, and the number of possible configurations where biases and covariates appear can be quite numerous in general. Mapping out the theoretical and practical identifiability requirements for a given SSM will remain an ongoing topic of research. 

\section*{Acknowledgements} The California Department of Water Resources and the Interagency Ecological Program for the San Francisco Estuary provided funding for this work. The findings and conclusions in this article are those of the authors and do not necessarily represent the view of the member agencies of the Interagency Ecological Program for the San Francisco Estuary. We thank William Smith, Matt Nobriga, William Bennett, Wim Kimmerer, Pete Smith, Randall Baxter, Fred Feyrer, Perry de Valpine, David Hankin, Wendy Meiring, Emilio Laca, Mark Maunder, Ruth King, and Diana Cole for helpful discussions.

\section*{Supporting Information}
Additional supporting information may be found online in the Supporting Information section at the end of the article.

\clearpage
\newpage

\appendix

\begin{center}
\large{Supporting Information for ``Improving inference for nonlinear state-space models of animal population dynamics given biased sequential life stage data''\\by\\ L. Polansky, K. B. Newman, and L. Mitchell}    
\end{center}

\renewcommand\thesection{Appendix \Alph{section}}
\renewcommand*\thefigure{\Alph{section}.\arabic{figure}}
\renewcommand*\thetable{\Alph{section}.\arabic{table}}
\renewcommand{\theequation}{\Alph{section}.\arabic{equation}}

\ssection{\\Analytical derivation of parameter identifiability}\label{app:model-identifiability}
\setcounter{equation}{0}
This section applies the methods of \citet{Cole-McCrea-2016}, and in particular following their Web Appendix A.4, to show that the mean and variance parameters  of the state-space model described in Section~\ref{sec:model-description} are estimable. For simplicity, only one covariate per vital rate is assumed.  Furthermore, while external,
and time-varying, estimates of observation variance are available for the case study, we considered the situation where such estimates were not available but the observation variances were time invariant.  The resulting parameter vector has 18 components: $\theta$=($\zeta_{0}$, $\zeta_{1}$, $\beta_{0}$, $\beta_{1}$,
$\eta_{0}$, $\eta_{1}$, $\gamma_{0}$, $\gamma_{1}$, $\psi_{J}$, $\psi_{SA}$,
 $\sigma_{P,R}^2$, $\sigma_{P,PL}^2$, $\sigma_{P,J}^2$, $\sigma_{P,SA}^2$,
 $\sigma^2_{O,PL}$, $\sigma^2_{O,J}$, $\sigma^2_{O,SA}$, $\sigma^2_{O,A}$).
 
  In the following we partition the calculation of the exhaustive summary
  into three pieces, one for the mean parameters of the state and observation
  models (namely the covariate coefficients and the observation bias
  parameters), another for the variance parameters of the state model,
  and another for the variance parameters of the observation model.
  We note that the initial analysis with \texttt{Maple} proceeded sequentially,
  starting with the first piece, then adding the second piece, and
  finally adding the third piece.  In each case, all parameters
  were separately identifiable at each stage (10, then 14, and then 18
  parameters).

\paragraph{Exhaustive summaries of covariate coefficients and observation biases.}
For simplicity we assumed just a single covariate per process, as
the exhaustive summary vector would simply need to be extended
in the case of more covariates, and assume that $n_{A,0}$ is known.
Given an unbiased estimate
of $n_{A,0}$ (namely, 1994 adults), $n_{A,0}$ is identifiable.  That leaves 10 unknown parameters. Thus at least 10 components to the exhaustive summary are needed. However, based on preliminary 
analyses with \texttt{Maple}, the ``first'' 12 observations were needed.
 
The lognormal bias corrections in the observations simplify
the expected values of the observations (compared to non-bias
corrections); e.g. $E[y_{PL,1}]$ = $E[E[y_{PL,1}|n_{PL,1}]]$ = $E[n_{PL,1}]$.
\begin{align*}
 E[y_{PL,1}] &= E[n_{PL,1}] \approx e^{\zeta_0+\zeta_1 x_{R,1}}n_{A,0}\\
 E[y_{J,1}] &= E[\psi_J n_{J,1}] \approx 
 \psi_J \frac{e^{\beta_0 + \beta_1 x_{PL,1}}}
 {1+e^{\beta_0 + \beta_1 x_{PL,1}}} e^{\zeta_0+\zeta_1 x_{R,1}}n_{A,0} \\
 E[y_{SA,1}] &= E[\psi_{SA} n_{SA,1}] \approx 
 \psi_{SA} \frac{e^{\eta_0 + \eta_1 x_{J,1}}}
 {1+e^{\eta_0 + \eta_1 x_{J,1}}}
 \frac{e^{\beta_0 + \beta_1 x_{PL,1}}}
 {1+e^{\beta_0 + \beta_1 x_{PL,1}}} e^{\zeta_0+\zeta_1 x_{R,1}}n_{A,0} \\
 E[y_{A,1}] &= E[ n_{A,1}] \approx 
 \frac{e^{\gamma_0 + \gamma_1 x_{SA,1}}}
 {1+e^{\gamma_0 + \gamma_1 x_{SA,1}}}
 \frac{e^{\eta_0 + \eta_1 x_{J,1}}}
 {1+e^{\eta_0 + \eta_1 x_{J,1}}}
 \frac{e^{\beta_0 + \beta_1 x_{PL,1}}}
 {1+e^{\beta_0 + \beta_1 x_{PL,1}}} e^{\zeta_0+\zeta_1 x_{R,1}}n_{A,0} \\
 E[y_{PL,2}] &= E[ n_{A,2}] \approx 
e^{\zeta_0+\zeta_1 x_{R,2}}
 \frac{e^{\gamma_0 + \gamma_1 x_{SA,1}}}
 {1+e^{\gamma_0 + \gamma_1 x_{SA,1}}}
 \frac{e^{\eta_0 + \eta_1 x_{J,1}}}
 {1+e^{\eta_0 + \eta_1 x_{J,1}}}
 \frac{e^{\beta_0 + \beta_1 x_{PL,1}}}
 {1+e^{\beta_0 + \beta_1 x_{PL,1}}} e^{\zeta_0+\zeta_1 x_{R,1}}n_{A,0}  \\
 & \vdots \\
 E[y_{A,3}] &= \ldots 
\end{align*}
 where the expectations are approximated by the deterministic version of the model. 

Letting $z_j$ = $\exp(\zeta_0+\zeta_1 x_{R,j})$, 
$b_j$ = $expit(\beta_0 + \beta_1 x_{PL,j})$, 
$e_j$ = $expit(\eta_0 + \eta_1 x_{J,j})$, and
$g_j$ = $expit(\gamma_0 + \gamma_1 x_{SA,j})$, a
length 12 exhaustive summary can be written as follows.
\begin{align}
\label{eq:1st.term}
    E[y_{PL,1}] & \approx z_1 n_{A,0} \\
  E[y_{J,1}] &\approx \psi_J b_1 z_1 n_{A,0} \\  
E[y_{SA,1}] & \approx \psi_{SA} e_1 b_1 z_1  n_{A,0} \\
 E[y_{A,1}] &  \approx g_1 e_1 b_1 z_1 n_{A,0} \\
 E[y_{PL,2}] & \approx z_2 g_1 e_1 b_1 z_1 n_{A,0} \\
 E[y_{J,2}] & \approx \psi_J b_2 z_2 g_1 e_1 b_1 z_1 n_{A,0} \\
 E[y_{SA,2}] & \approx \psi_{SA} e_2 b_2 z_2 g_1 e_1 b_1 z_1 n_{A,0} \\
 E[y_{A,2}] & \approx  g_2 e_2 b_2 z_2 g_1 e_1 b_1 z_1 n_{A,0} \\
 E[y_{PL,3}] & \approx  z_3 g_2 e_2 b_2 z_2 g_1 e_1 b_1 z_1 n_{A,0} \\
 E[y_{J,3}] & \approx  \psi_J b_3 z_3 g_2 e_2 b_2 z_2 g_1 e_1 b_1 z_1 n_{A,0}  \\
 E[y_{SA,3}] & \approx  \psi_{SA} e_3 b_3 z_3 g_2 e_2 b_2 z_2 g_1 e_1 b_1 z_1 n_{A,0}  \\
 \label{eq:12th.term}
 E[y_{A,3}] & \approx  g_3 e_3 b_3 z_3 g_2 e_2 b_2 z_2 g_1 e_1 b_1 z_1 n_{A,0}  
\end{align}

\paragraph{Exhaustive summary for
state process variance parameters.}
Following \citet{Cole-McCrea-2016} (based on their Web Appendix A.4), to determine the identifiability
of the variance parameters  the exhaustive summary
is expanded to include \textit{unconditional} variances of observations. 

Two general results that are used repeatedly to  approximate 
the unconditional variances of the observations are:
 \begin{enumerate}

    \item For two independent random variables $X$ and $Y$:
\begin{align*}
    V[XY] &= V[X]E[Y]^2 + E[X]^2V[Y] + V[X]V[Y]
\end{align*}
 \item $V[Y]$ = $E[V(Y|X)]$ + $V[E(Y|X)]$. In the application
 here $Y$ is the observation and $X$ is the underlying state
 or the product of an observation bias parameter ($\psi_J$
 or $\psi_{SA}$) and the state. For example,
 \begin{align*}
     V[y_{PL,1}] &= E[V(y_{PL,1}|n_{PL,1})] + V[E(y_{PL,1}|n_{PL,1})]
     \\
    &= \sigma^2_{O,PL,1} + V[n_{PL,1}]
 \end{align*}
\end{enumerate}
 
Focusing on exhaustive summaries for the four process variances,
the variances of the first four observations are 
considered:
\begin{align}
\label{eq:13th.term}
      V[y_{PL,1}] &= \sigma^2_{O,PL,1} + V[n_{PL,1}] = \sigma^2_{O,PL,1} + V[\rho_1 n_{A,0}]  =
      \sigma^2_{O,PL,1} +  n_{A,0}^2 V[\rho_1]
       \\
    V[y_{J,1}]  &= \sigma^2_{O,J,1} + V[\psi_J n_{J,1}] =
    \sigma^2_{O,J,1} + (\psi_J n_{A,0})^2 V[\phi_{PL,1}*\rho_1]
   \\ 
    V[y_{SA,1}] &= \sigma^2_{O,SA,1} + V[\psi_{SA} n_{SA,1}] =
     \sigma^2_{O,SA,1} +(\psi_{SA}  n_{A,0})^2 
        V[\phi_{J,1} * \phi_{PL,1} * \rho_1]\\
\label{eq:16th.term}
    V[y_{A,1}]  &= \sigma^2_{O,A,1} + V[n_{A,1}] =
    \sigma^2_{O,A,1} + n_{A,0}^2 
    V[\phi_{SA,1}* \phi_{J,1} * \phi_{PL,1} * \rho_1]
\end{align}
where, letting 
\begin{align*}
     s_1     &= e^{\sigma^2_{P,R}}-1 \\ 
     s_{2,1} &=   \frac{\sigma^2_{P,PL}}
                {(1+e^{\beta_0+\beta_1 x_{PL,1}})^2}     \\
s_{3,1} &=   \frac{\sigma^2_{P,J}}
                  {(1+e^{\eta_0+\eta_1 x_{J,1}})^2}           \\
     s_{4,1} &=   \frac{\sigma^2_{P,SA}}
                             {(1+e^{\gamma_0+\gamma_1 x_{SA,1}})^2}
\end{align*} 

\begin{align*}
 V[\rho_1] &= z_1^2 s_1 \\
 V[\phi_{PL,1}*\rho_1] & 
  \approx b_1^2 z_1^2(s_{2,1}  + s_1 + s_{2,1}   s_1 )\\
 V[\phi_{J,1} * \phi_{PL,1} * \rho_1] &\approx 
 e_1^2 b_1^2 z_1^2 \left [ s_{3,1}(1+s_{2,1}+s_1+s_{2,1}s_1) +
   (s_{2,1}+s_1+s_{2,1}s_1) \right ]\\
 V[\phi_{SA,1}* \phi_{J,1} * \phi_{PL,1} * \rho_1] &\approx 
 g_1^2e_1^2b_1^2z_1^2\left [s_{4,1} + (1+s_{4,1})
 [s_{3,1}(1+s_{2,1}+s_1+s_{2,1}s_1) + (s_{2,1}+s_1 + s_{2,1}s_1)] \right ] 
\end{align*}
 where the  variance for $\rho_1$, which is based on a lognormal random variable, is exact but the remaining variances are approximations of a logit-normal variance calculated using the delta method.

\paragraph{Exhaustive summary for
observation variance parameters.}
We assume that observation variances are life-stage, but not time,
specific, and the exhaustive summary is extended
with four more components.  This is
done by calculating the variances of the next four observations,
namely, $y_{PL,2}$, $y_{J,2}$, $y_{SA,2}$, and $y_{A,2}$.
\begin{align}
\label{eq:17th.term}
      V[y_{PL,2}] &= \sigma^2_{o,PL} + V[n_{PL,2}] =  
      \sigma^2_{o,PL} +  n_{A,0}^2 V[\rho_2~ \phi_{SA,1}~\phi_{J,1}  ~\phi_{PL,1}~\rho_1]
       \\
    V[y_{J,2}]  &= \sigma^2_{o,J} + V[\psi_J n_{J,2}] =
    \sigma^2_{o,J} + (\psi_J n_{A,0})^2 
    V[\phi_{PL,2}~\rho_2~\phi_{SA,1}~\phi_{J,1}~\phi_{PL,1}~\rho_1]
   \\ 
    V[y_{SA,2}] &= \sigma^2_{o,SA} + V[\psi_{SA} n_{SA,2}] =
     \sigma^2_{o,SA} +(\psi_{SA}  n_{A,0})^2 
        V[\phi_{J,2}~\phi_{PL,2}~\rho_2~\phi_{SA,1}~ \phi_{J,1}~\phi_{PL,1}~\rho_1]\\
\label{eq:20th.term}
    V[y_{A,2}]  &= \sigma^2_{o,A} + V[n_{A,2}] =
    \sigma^2_{o,A} + n_{A,0}^2 
    V[\phi_{SA,2}~\phi_{J,2}~\phi_{PL,2}~\rho_2 \phi_{SA,1}~\phi_{J,1}~\phi_{PL,1}~\rho_1] 
\end{align}
where
{\scriptsize 
\begin{align*}
 V[\rho_2~\phi_{SA,1}~\phi_{J,1}~\phi_{PL,1}~\rho_1] &\approx 
 V[\rho_2](g_1^2e_1^2b_1^2z_1^2)+
 (V[\rho_2]+E[\rho_2]^2)V[\phi_{SA,1} \phi_{J,1}  \phi_{PL,1}  \rho_1] \\
 &= z_2^2 s_1(g_1^2e_1^2b_1^2z_1^2) + (z_2^2 s_1 + z_2^2)V[\phi_{SA,1} \phi_{J,1}  \phi_{PL,1}  \rho_1] \\
V[\phi_{PL,2}~\rho_2~\phi_{SA,1}~\phi_{J,1}~\phi_{PL,1}~\rho_1]
&\approx b_2^2s_{2,2}(z_2^2g_1^2e_1^2b_1^2z_1^2) +
(b_2^2s_{2,2}+b_2^2)V[\rho_2~\phi_{SA,1}~\phi_{J,1}~\phi_{PL,1}~\rho_1] \\
V[\phi_{J,2}~\phi_{PL,2}~\rho_2~\phi_{SA,1}~\phi_{J,1}~ \phi_{PL,1}~\rho_1] &\approx e_2^2s_{3,2}(b_2^2z_2^2g_1^2e_1^2b_1^2z_1^2) +
 (e_2^2s_{3,2}+e_2^2)V[\phi_{PL,2}~\rho_2~\phi_{SA,1}~\phi_{J,1}~\phi_{PL,1}~\rho_1] \\
V[\phi_{SA,2}~\phi_{J,2}~\phi_{PL,2}~\rho_2 \phi_{SA,1}~\phi_{J,1}~\phi_{PL,1}~\rho_1] &\approx
g_2^2s_{4,2}(e_2^2b_2^2z_2^2g_1^2e_1^2b_1^2z_1^2) +
(g_2^2s_{4,2}+g_2^2)V[\phi_{J,2}~\phi_{PL,1}~\phi_{PL,2}~\rho_2~\phi_{SA,1}~\phi_{J,1}~ \phi_{PL,1}~\rho_1]
\end{align*}}
where $s_{2,2}$, $s_{3,2}$, and $s_{4,2}$ are analogous to $s_{2,1}$,
$s_{3,1}$, and $s_{4,1}$ but using the covariates for time 2.

\paragraph{Results.}The exhaustive summary vector for the 18 parameters consisted of the
expressions from the three blocks of equations \ref{eq:1st.term}-\ref{eq:12th.term}, \ref{eq:13th.term}-\ref{eq:16th.term}, and \ref{eq:17th.term}-\ref{eq:20th.term}. The derivative matrix, $D$, 
of dimension 18 by 20, which was symbolically
calculated using \texttt{Maple}, had rank 18,   indicating
that all 18 parameters are identifiable.

\newpage 

\ssection{\\Simulation study details}\label{app:sim-study}
\setcounter{figure}{0}
\setcounter{table}{0}

\begin{table}[ht]
\centering
\caption{Simulation input values and prior distributions.}
\begin{tabular}{ l l r l}
  \hline \hline
Description & Parameter & True Value & Prior Distribution \\ 
  \hline
  Initial abundance & $n_{A,0}$ & 100,000 & $\text{Unif}(10000, 200000)$ \\
  Recruitment & $\zeta_0$ & 1 & $\text{Norm}(0, 0.1)$ \\
              & $\zeta_1$ & 1 & $\text{Norm}(0, 0.1)$ \\
              & $\zeta_2$ & 0 & $\text{Norm}(0, 0.1)$ \\
              & $\sigma_{P,R}$ & 0.05 & $\text{Exponential}(\text{rate}=5)$ \\
  Survival & &  \\
  $\phi_{PL}$ & $\beta_{0,1}$ & 1.3 & $\text{Norm}(0, \sqrt{\pi^2/9})$ \\
           & $\beta_{1,1}$ & 1 & $\text{Norm}(0, \sqrt{\pi^2/9})$ \\
           & $\beta_{2,1}$ & 0 & $\text{Norm}(0, \sqrt{\pi^2/9})$ \\
           & $\sigma_{P,PL}$ & 0.5 & $\text{Exponential}(\text{rate}=0.75)$ \\
  $\phi_{J}$ & $\beta_{0,2}$ & 1.3 & $\text{Norm}(0, \sqrt{\pi^2/9})$ \\
           & $\beta_{1,2}$ & 1 & $\text{Norm}(0, \sqrt{\pi^2/9})$ \\
           & $\beta_{2,2}$ & 0 & $\text{Norm}(0, \sqrt{\pi^2/9})$ \\
           & $\sigma_{P,J}$ & 0.5 & $\text{Exponential}(\text{rate}=0.75)$ \\ 
  $\phi_{SA}$ & $\beta_{0,3}$ & 1.3 & $\text{Norm}(0, \sqrt{\pi^2/9})$ \\
           & $\beta_{1,3}$ & 1 & $\text{Norm}(0, \sqrt{\pi^2/9})$ \\
           & $\beta_{2,3}$ & 0 & $\text{Norm}(0, \sqrt{\pi^2/9})$ \\
           & $\sigma_{P,SA}$ & 0.5 & $\text{Exponential}(\text{rate}=0.75)$ \\ 
  Observation measurement bias & &  \\
           & $\psi_{J}$ & 0.5 & $\text{Exponential}(\text{rate}=1)$ \\ 
           & $\psi_{SA}$ & 0.2 & $\text{Exponential}(\text{rate}=1)$ \\
  \multirow{2}{5cm}{Observation measurement coefficient of variation} & &  \\
           & $\widehat{CV}[\hat{n}_{s,t}]_{Ex}$ & $\text{Uniform}(0.1,1)$ & $\text{Exponential}(\text{rate}=1)$ \\ 
   \hline
\end{tabular}
\label{tab:sim.true.value.table}
\end{table}

\break 
\newpage
\ssection{\\Case study details}\label{app:case-study}
\setcounter{figure}{0}
\setcounter{table}{0}
Covariate data was obtained from a variety of California State and U.S. Federal government agencies. Subsets of the data from areas of the Delta most likely to coincide with delta smelt presence were then temporally summarized using means or medians. Table~\ref{tab:predictor-variables} summarizes the types of data considered. For exploratory and management purposes, more predictor variables for any given vital rate were constructed than could be simultaneously included in a global model because of collinearity (e.g. spring inflow, outflow, and OMR, or summer X2 and NJACM).

Prior to fitting models which included multiple covariates per vital rate (''global'' models), we fit a series of relatively simple models to examine the influence of a single type of covariate on each of the vital rates by graphical examination of vital rate predictions. These models included a single predictor variable for each vital rate where each was of the same type, e.g. an all inflow model used spring, summer, fall and winter measures of inflow to predict each of the corresponding vital rates. This was helpful for identifying the strength of a relationship for each covariate alone when including other ones measuring similar features of the ecosystem would be difficulty because of collinearity. The results also give a preliminary indication about whether any vital rates show a response in the direction opposite that expected and thus warrant further consideration of inclusion based on biological reasonableness. 

Given a fitted model, vital rate prediction intervals that included posterior uncertainty were constructed by first sampling a vector of parameters from the joint posterior distribution of the fitted model, and then simulating a realized vital rate given this sample, repeated 100,000 times. To evaluate the influence of posterior uncertainty on the prediction interval, the expected vital rate values as a function of the covariate given the mean posterior values of the controlling parameters was also computed. 

\begin{figure}
    \centering
    \includegraphics[scale=1]{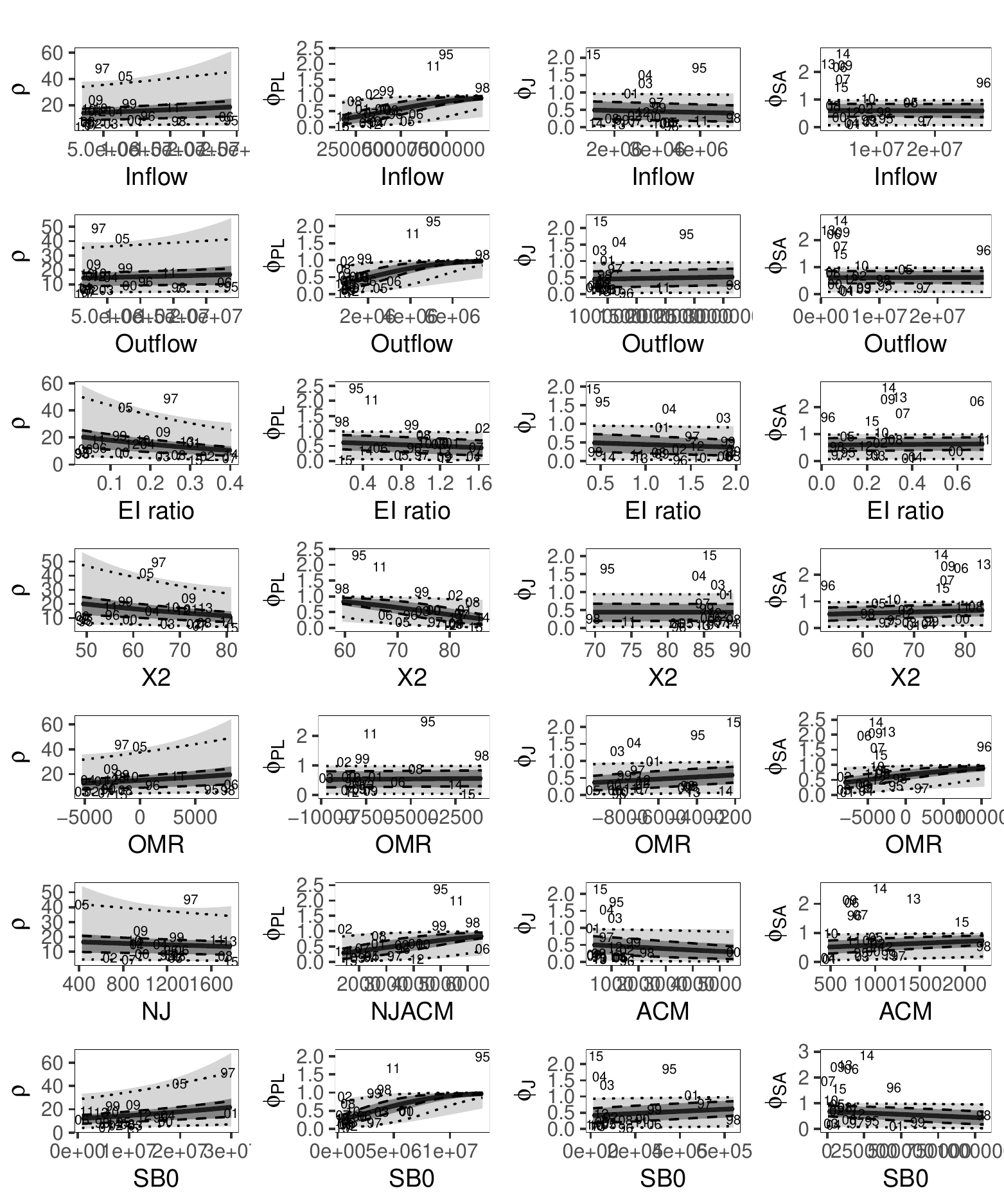}
    \caption{Predicted vital rates for preliminary focal models using a single predictor variable of the same type for each state process equation. Each panel row corresponds to different model and the columns are the different vital rates. The solid curved lines show expected values, dark and light grey shadings show the $100(1-\alpha)\%$ central credible intervals for $\alpha=0.5$ and $\alpha=0.05$, respectively, and include posterior parameter estimate uncertainty. The dashed and dotted lines show the 50\% and 95\%, respectively, central credible intervals using the mean values of the posterior.}
    \label{fig:set-of-focal-models}
\end{figure}

Figure~\ref{fig:set-of-focal-models} shows the predicted vital rates for the models with a single covariate per vital rate. In general, the expected relationship between the covariate and the vital rate held. A notable exception was the model using early life stage striped bass (SB0), which can only be detrimental to delta smelt recruitment and survival, but for which the modeled recruitment $\rho$, and summer ($\phi_{PL}$) and fall ($\phi_{J}$) survival rates responded positively to increases in their densities. This suggests that the decline in these conspecifics along with delta smelt based on shared habitat quality during the time period studied overwhelms any direct interspecific interaction effects. Based on this, the SB0 covariate was not considered in the global models. Other predator/competitor indices were retained in the global models without prior inspection of the direction of their effects because they were involved in too few vital rates to allow model estimation alone.

\begin{table}
\centering
\caption{Covariates considered for modeling recruitment and survival. For each covariate, the vital rates it was used for are shown in the vital rates column. Generally covariates are means or medians over Mar-May, Jun-Aug, Sep-Nov, and Dec-Feb for $\rho$, $\phi_{PL}$, $\phi_{J}$, and $\phi_{SA}$, with a few exceptions (that shift the start and end times by one month) related to data availability or management needs. The direction column shows the expected effect. The \textsuperscript{*} indicates the subset of covariates considered in the global model after removing highly collinear covariates or those that had strong effects in the opposite of their expected direction.}
\label{tab:predictor-variables}
\begin{tabular}{lllp{8.5cm}}
  \hline\hline
 \multicolumn{1}{c}{Predictor} &  \multicolumn{1}{c}{Vital rates} & Direction & \multicolumn{1}{c}{Remarks} \\ 
  \hline
Inflow & $\rho^{*}$, $\phi_{PL}^{*}$, $\phi_{J}$, $\phi_{SA}^{*}$ & $+$ & Inflow, an omnibus habitat condition measure.\\
Outflow & $\rho$, $\phi_{PL}$, $\phi_{J}$, $\phi_{SA}$  & $+$ & Outflow, an omnibus habitat condition measure.\\
EI ratio & $\rho$, $\phi_{PL}$, $\phi_{J}$, $\phi_{SA}$  & $-$ & Total exports to total inflow ratio.\\
OMR & $\rho$, $\phi_{PL}$, $\phi_{J}$, $\phi_{SA}^{*}$ & $+$ & Old and Middle river flows.\\
X2 & $\rho^{*}$, $\phi_{PL}$, $\phi_{J}^{*}$, $\phi_{SA}$  & $-$ & Location of the 2\% salinity isocline location. $\textrm{X2}_{t-1}$ denotes the fall X2 location of the prior calendar year used for making predictions about cohort $t$ recruitment.\\
LSZ & $\phi_{PL}$, $\phi_{J}$ & $+$ & Volume of low salinty zone habitat.\\
Mallard & $\phi_{PL},\phi_{J}$ & $-$ & Salinity at Mallard Island.\\
Secchi & $\rho^{*}$, $\phi_{PL}^{*}$, $\phi_{J}^{*}$, $\phi_{SA}^{*}$  & $-$ & Water clarity as measured by Secchi depth (cm). South Secchi is a covariate constructed from Secchi data only from the South portion of the Delta.\\
Temperature & $\rho^{*}$, $\phi_{PL}^{*}$, $\phi_{J}^{*}$, $\phi_{SA}^{*}$  & $-/unc$ & Temperature (deg C). Cooler temperatures are expected to be better for recruitment and summer survival. Temperature effects for fall and winter survival are uncertain (unc).\\
ACM & $\rho^{*}$, $\phi_{J}^{*}$, $\phi_{SA}^{*}$  & $+$ & Large prey availability for late juveniles and adults.\\
NJ & $\rho^{*}$  & $+$ & Small prey availability for early life history fish.\\
NJACM & $\phi_{PL}$  & $+$ & Combined small and large prey availability. Highly correlated with inflow in the summer.\\
ISS& $\rho^{*}$, $\phi_{PL}^{*}$  & $-$ & Inland silverside \textit{Menidia beryllina}, a competitor.\\
SB0 & $\rho$, $\phi_{PL}$, $\phi_{J}$, $\phi_{SA}$  & $-$ & Juvenile striped bass \textit{Morone saxatilis}, a competitor.\\
SB1 & $\phi_{SA}^{*}$  & $-$ & Adult striped bass, a predator.\\
TFS & $\rho^{*}$, $\phi_{PL}^{*}$  & $-$ & Threadfin shad \textit{Dorosoma petenense}, a competitor.\\
TG & $\rho^{*}$, $\phi_{PL}^{*}$  & $-$ & Tridentiger goby \textit{Tridentiger spp.}, a competitor.\\
   \hline
\end{tabular}
\end{table}

\break 
\newpage

\begin{figure}[h]
    \centering
    \includegraphics[scale=1]{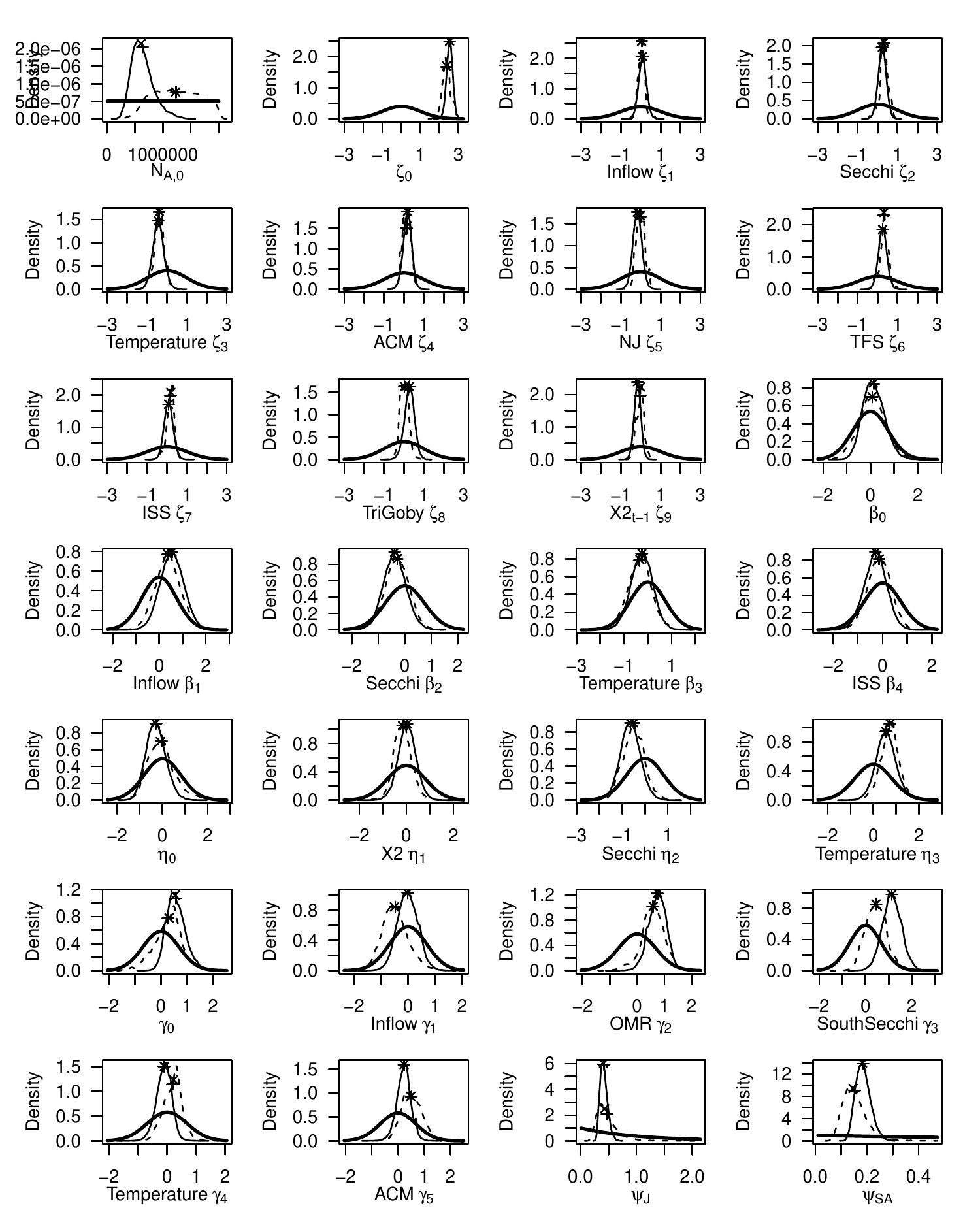}
    \caption{Prior and posterior distributions for the initial latent abundance, state process parameters, and observation bias parameters. Thick line- prior; thin line- posterior from model using externally calculated observation error CV; dashed line- posterior from model internally estimating observation error CV; + is at the mean; x is at the median.}
    \label{fig:Global_models_prior_vs_posterior}
\end{figure}

\begin{figure}[h]
    \centering
    \includegraphics[scale=1]{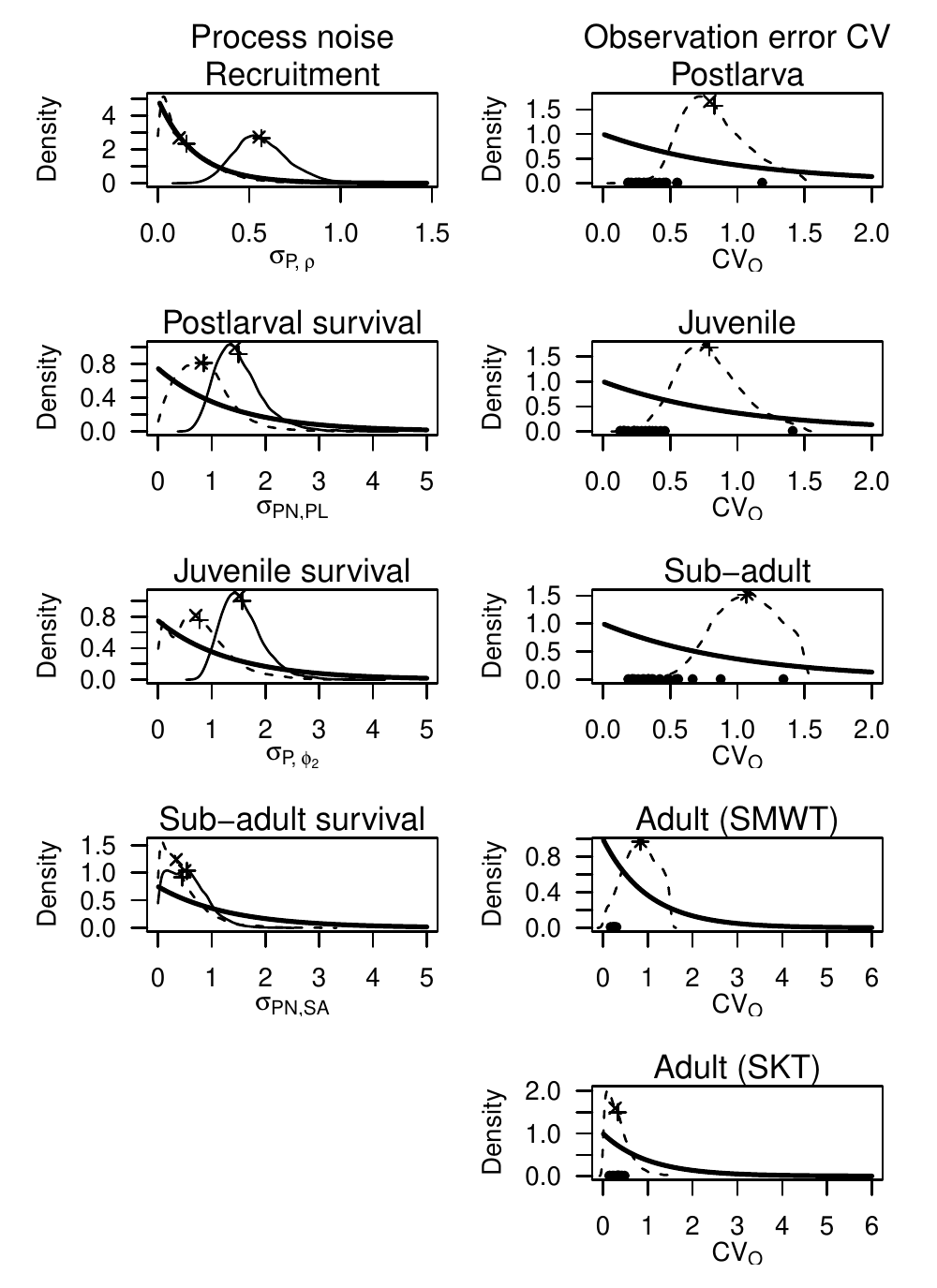}
    \caption{Variance parameter comparisons of prior (solid thick line) and posterior distributions. Process variance (left column) and observation error coefficient of variation (right column) from the delta smelt global models.  Thick line- prior; thin line- posterior from model using externally calculated observation error CV; dashed line- posterior from model internally estimating observation error CV; + is at the mean; x is at the median; dots in the left column along the x-axis are the externally (to the SSM) estimated values.}
    \label{fig:Global_model_PN_OE_variances}
\end{figure}

\begin{figure}[h]
    \centering
    \includegraphics[scale=1]{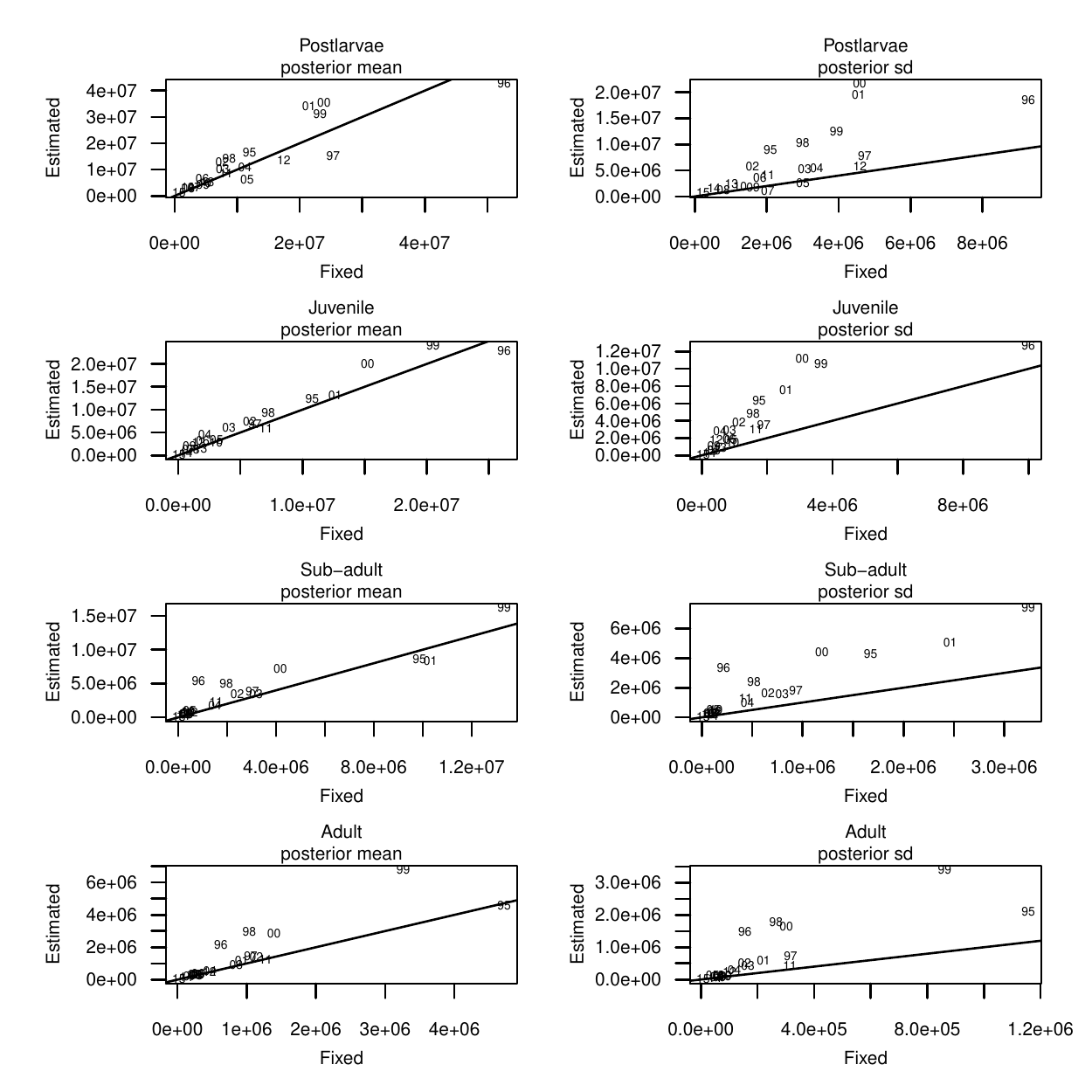}
    \caption{Latent abundance posterior means (left column) and standard deviations (right column) based on the global model fitted with fixed observation error CV (x-axis) and estimated  observation error CV (y-axis). Numbers in the plot indicate the cohort.}
    \label{fig:Global_models_latent_abund_mean_and_sd}
\end{figure}

\begin{figure}[h]
    \centering
    \includegraphics[scale=1]{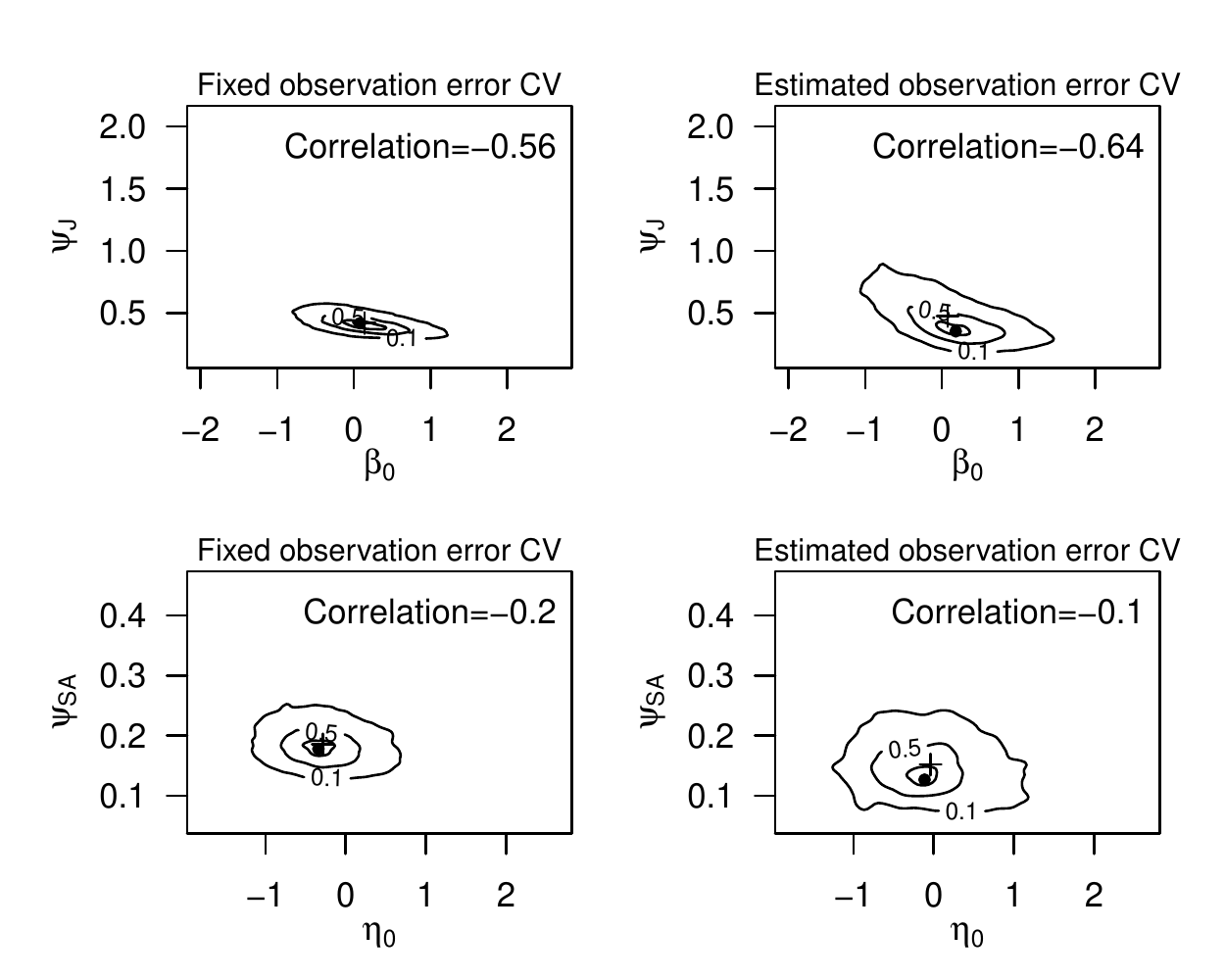}
    \caption{Joint posterior density plots of the survival intercept and  observation bias parameters for juvenile (top row) and sub-adult (bottom row) survival and observation models. Joint denisty plots are scaled to lie between 0 and 1, levels are drawn at the 0.9, 0.5 and 0.1 quantiles of the distribution, the dot is at the maximum, and the cross is at the marginal means.}
    \label{fig:case_study_psi_vs_intercepts}
\end{figure}

Using the subset of covariates indicated in Table~\ref{tab:predictor-variables}, two global models including these covariates to construct multivariate models for each vital rater were considered: one that used external estimates of observation error CV, and one that internally estimated obseration error CV along with the observation error bias and the state process parameters. Table~\ref{tab:post-summary-fix-vs-est} summarizes the (non-latent state) parameter marginal posterior for the global models, and Figures ~\ref{fig:Global_models_prior_vs_posterior} and \ref{fig:Global_model_PN_OE_variances} show the corresponding prior and posterior distributions. Figure~\ref{fig:Global_models_latent_abund_mean_and_sd} compares posterior means and standard deviations of latent abundance estimates between the global model fit with fixed observation error CV and the global model with estimated observation error CV. Figure~\ref{fig:case_study_psi_vs_intercepts} illustrates how diffusivity increases for bias parameters $\psi$ and vital rate intercept parameters; compare with simulation study Figure~\ref{fig:sim.results.psi.intercept.jposteriors}. 

\begin{table}
\centering
\caption{Posterior mean, standard deviation (Std), evidence (Evi), and lower and upper 95\% highest posterior density intervals of parameters for the global model with either fixed or estimated observation error CV. Evidence is the proportion of the posterior above zero when the expected effect is positive (Direction=pos), proportion of the posterior below zero when the expected effect is negative (Direction=neg); if Direction is uncertain (unc), evidence is the proportion above or below zero if the mean is above or below zero, respectively.}
\label{tab:post-summary-fix-vs-est}
\begin{tabular}{llrrrrrrrrrr}
  \hline\hline
  &&\multicolumn{5}{c}{Fixed observation error CV}&\multicolumn{5}{c}{Estimated observation error CV}\\
\cmidrule(lr){3-7}
\cmidrule(lr){8-12}
Parameter & Dir & Mean & Std & Evi & Lower & Upper & Mean & Std & Evi & Lower & Upper\\ 
  \hline
    && \multicolumn{10}{c}{Recruitment}\\
    \cline{2-12}
Intercept &  & 2.54 & 0.16 &  & 2.22 & 2.86 & 2.38 & 0.23 &  & 1.98 & 2.87 \\ 
 Inflow & pos & 0.09 & 0.20 & 0.67 & -0.31 & 0.50 & 0.05 & 0.18 & 0.60 & -0.30 & 0.43 \\ 
  Secchi & neg & 0.23 & 0.21 & 0.13 & -0.18 & 0.64 & 0.32 & 0.23 & 0.09 & -0.17 & 0.77 \\ 
  Temperature & neg & -0.44 & 0.28 & 0.94 & -0.98 & 0.13 & -0.40 & 0.26 & 0.94 & -0.96 & 0.10 \\ 
  ACM & pos & 0.17 & 0.22 & 0.80 & -0.27 & 0.59 & 0.14 & 0.21 & 0.70 & -0.22 & 0.53 \\ 
  NJ & pos & -0.17 & 0.23 & 0.23 & -0.61 & 0.29 & -0.01 & 0.22 & 0.46 & -0.37 & 0.46 \\ 
  TFS & neg & 0.25 & 0.22 & 0.12 & -0.18 & 0.69 & 0.30 & 0.23 & 0.12 & -0.20 & 0.70 \\ 
  ISS & neg & 0.08 & 0.23 & 0.37 & -0.36 & 0.55 & 0.17 & 0.20 & 0.17 & -0.30 & 0.51 \\ 
  TG & neg & 0.27 & 0.26 & 0.14 & -0.25 & 0.76 & 0.03 & 0.22 & 0.46 & -0.41 & 0.43 \\ 
  $\textrm{X2}_{t-1}$ & neg & -0.16 & 0.18 & 0.83 & -0.51 & 0.18 & -0.03 & 0.19 & 0.53 & -0.41 & 0.33 \\ 
  && \multicolumn{10}{c}{Postlarval survival}\\
\cline{2-12}
Intercept &  & 0.14 & 0.46 &  & -0.71 & 1.06 & 0.08 & 0.57 &  & -1.07 & 1.20 \\  
Inflow & pos & 0.55 & 0.51 & 0.86 & -0.44 & 1.53 & 0.36 & 0.54 & 0.75 & -0.71 & 1.37 \\ 
  Secchi & neg & -0.38 & 0.44 & 0.81 & -1.26 & 0.47 & -0.29 & 0.46 & 0.74 & -1.22 & 0.57 \\ 
  Temperature & neg & -0.25 & 0.46 & 0.71 & -1.20 & 0.62 & -0.35 & 0.51 & 0.76 & -1.42 & 0.63 \\ 
  ISS & neg & -0.29 & 0.46 & 0.74 & -1.17 & 0.63 & -0.14 & 0.48 & 0.61 & -1.07 & 0.81 \\ 
  TG & neg & 0.11 & 0.49 & 0.41 & -0.83 & 1.07 & -0.28 & 0.46 & 0.73 & -1.20 & 0.62 \\ 
   && \multicolumn{10}{c}{Juvenile survival}\\
\cline{2-12}
Intercept &  & -0.28 & 0.44 &  & -1.13 & 0.56 & -0.04 & 0.62 &  & -1.11 & 1.36 \\ 
X2 & neg & 0.03 & 0.39 & 0.47 & -0.75 & 0.78 & -0.16 & 0.39 & 0.68 & -0.91 & 0.64 \\ 
  Secchi & neg & -0.64 & 0.44 & 0.93 & -1.52 & 0.24 & -0.49 & 0.47 & 0.87 & -1.44 & 0.39 \\ 
  Temperature & unc & 0.56 & 0.45 & 0.90 & -0.31 & 1.46 & 0.72 & 0.42 & 0.95 & -0.14 & 1.55 \\ 
  ACM & pos & -0.27 & 0.37 & 0.23 & -1.00 & 0.47 & 0.03 & 0.40 & 0.49 & -0.70 & 0.88 \\ 
 && \multicolumn{10}{c}{Sub-adult survival}\\
\cline{2-12}
Intercept &  & 0.57 & 0.36 &  & -0.08 & 1.31 & 0.26 & 0.52 &  & -0.83 & 1.24 \\  
Inflow & pos & -0.02 & 0.38 & 0.48 & -0.74 & 0.74 & -0.48 & 0.49 & 0.15 & -1.39 & 0.54 \\ 
  OMR & pos & 0.75 & 0.33 & 0.99 & 0.09 & 1.36 & 0.58 & 0.42 & 0.92 & -0.25 & 1.43 \\ 
  South Secchi & pos & 1.14 & 0.41 & 1.00 & 0.34 & 1.97 & 0.46 & 0.46 & 0.84 & -0.41 & 1.35 \\ 
  Temperature & unc & -0.11 & 0.26 & 0.35 & -0.64 & 0.37 & 0.19 & 0.32 & 0.73 & -0.47 & 0.79 \\ 
  ACM & pos & 0.22 & 0.27 & 0.81 & -0.31 & 0.74 & 0.51 & 0.38 & 0.91 & -0.21 & 1.24 \\ 
  SB1 & neg & -0.23 & 0.24 & 0.86 & -0.72 & 0.26 & -0.05 & 0.37 & 0.59 & -0.73 & 0.77 \\ 
     && \multicolumn{10}{c}{Process variance}\\
\cline{2-12}
$\sigma_{P,R}$ & &  0.57 & 0.15 &  & 0.29 & 0.87 & 0.16 & 0.14 &  & 0.00 & 0.44 \\ 
$\sigma_{P,PL}$ & &  1.50 & 0.45 &  & 0.73 & 2.39 & 0.85 & 0.50 &  & 0.01 & 1.75 \\ 
$\sigma_{P,J}$ & &  1.57 & 0.40 &  & 0.87 & 2.38 & 0.78 & 0.53 &  & 0.00 & 1.79 \\ 
$\sigma_{P,SA}$ & &  0.54 & 0.36 &  & 0.00 & 1.20 & 0.45 & 0.41 &  & 0.00 & 1.27 \\
   && \multicolumn{10}{c}{Observation error bias}\\
\cline{2-12}
$\psi_{STN}$ & & 0.42 & 0.07 &  & 0.29 & 0.55 & 0.47 & 0.19 &  & 0.19 & 0.85 \\ 
$\psi_{FMWT}$ & & 0.19 & 0.03 &  & 0.13 & 0.24 & 0.15 & 0.05 &  & 0.07 & 0.24 \\    
&& \multicolumn{10}{c}{Observation error CV}\\
\cline{2-12}
$CV[n_{PL}]_{O,In}$ & & & & & & &  0.83 & 0.24 &  & 0.41 & 1.35 \\ 
$CV[n_{J}]_{O,In}$ & & & & &  & &0.79 & 0.24 &  & 0.36 & 1.30 \\  
$CV[n_{SA}]_{O,In}$ & & & & & & & 1.06 & 0.23 &  & 0.68 & 1.50 \\   
$CV[n_{A,SMWT}]_{O,In}$ & & & & & & & 0.83 & 0.36 &  & 0.22 & 1.50 \\ 
$CV[n_{A,SKT}]_{O,In}$ & & & & &  & & 0.33 & 0.26 &  & 0.01 & 0.83 \\ 
   \hline
\end{tabular}
\end{table}

\end{document}